\documentclass[a4paper,11pt]{article}
\usepackage{jheppub} 
\usepackage{lineno}
\usepackage{xcolor}
\usepackage{graphicx}
\usepackage{wrapfig}
\usepackage{amsmath}
\usepackage{float}
\usepackage{amssymb}
\usepackage{multirow}
\usepackage{slashed}
\usepackage{physics}
\usepackage{booktabs}
\usepackage[caption=false]{subfig}
\usepackage[colorlinks=true, linkcolor=blue, citecolor=blue, urlcolor=blue]{hyperref}
\usepackage{placeins}
\usepackage{hyperref}
\usepackage{orcidlink}

\newcommand \ms {m_s}

\newcommand \Schrodinger {Schr$\ddot{\mathrm{o}}$dinger }
\newcommand \meff {M_{\mathrm{eff}}}

\arxivnumber{2501.11257} 

\title{\boldmath In-medium bottomonium properties from lattice NRQCD calculations with extended meson operators}

\author[a]{Heng-Tong Ding\orcidlink{0000-0003-0590-081X},}
\author[a]{Wei-Ping Huang\orcidlink{0000-0003-0845-486X},}
\author[b]{Rasmus Larsen,}
\author[c]{Stefan Meinel\orcidlink{0000-0003-1034-1004},}
\author[d]{\\Swagato Mukherjee\orcidlink{0000-0002-3824-1008},}
\author[d]{Peter Petreczky\orcidlink{0000-0003-0285-2066}}
\author[e]{and Zhanduo Tang\orcidlink{0009-0003-3922-0334}}
\affiliation[a]{Key Laboratory of Quark and Lepton Physics (MOE) and Institute of Particle Physics, \\
Central China Normal University, Wuhan 430079, China}
\affiliation[b]{Fakult\"at f\"ur Physik, Universit\"at Bielefeld, \\
D-33615 Bielefeld, Germany}
\affiliation[c]{Department of Physics, University of Arizona, \\
Tucson, Arizona 85721, U.S.A.}
\affiliation[d]{Physics Department, Brookhaven National Laboratory, \\
Upton, New York 11973, U.S.A.}
\affiliation[e]{Cyclotron Institute and Department of Physics and Astronomy, Texas A\&M University, \\
College Station, TX 77843-3366, U.S.A.}

\emailAdd{hengtong.ding@ccnu.edu.cn}
\emailAdd{huangweiping@mails.ccnu.edu.cn}
\emailAdd{rlarsen@physik.uni-bielefeld.de}
\emailAdd{smeinel@arizona.edu}
\emailAdd{swagato@bnl.gov}
\emailAdd{petreczk@bnl.gov}
\emailAdd{zhanduotang@tamu.edu}

\abstract{We calculate the temperature dependence of bottomonium correlators in (2+1)-flavor lattice QCD with the aim to constrain in-medium properties of bottomonia at high temperature. The lattice calculations are performed using HISQ action with physical strange quark mass and light quark masses twenty times smaller than the strange quark mass at two lattice spacings $a=0.0493$ fm and $0.0602$ fm, and temporal extents $N_{\tau}=16-30$, corresponding to the temperatures $T=133-250$ MeV. We use a tadpole-improved NRQCD action including spin-dependent $v^6$ corrections for the heavy quarks and extended meson operators in order to be sensitive to in-medium properties of the bottomonium states of interest. We find that within estimated errors the bottomonium masses do not change compared to their vacuum values for all temperatures under our consideration; however, we find different nonzero widths for the various bottomonium states.}

\begin{document}
\maketitle
\flushbottom

\section{Introduction}
Matsui and Satz proposed quarkonium suppression can serve as a signature of quark-gluon plasma (QGP) formation in heavy-ion collisions~\cite{Matsui:1986dk}. Their proposal was based on the observation that color screening within a deconfined medium can make the interaction between the heavy quark and anti-quark short ranged, leading to the dissolution of quarkonia in QGP. Since then the study of quarkonium production in heavy-ion collisions has been a large part of the experimental heavy-ion program; see, e.g., Refs.~\cite{Aarts:2016hap,Zhao:2020jqu} for recent reviews.

The Matsui-Satz picture turned out to be incomplete. At nonzero temperature the potential is complex: in addition to the modification of its real part it also develops an imaginary part encoding the dissipative effects of the medium~\cite{Laine:2006ns,Brambilla:2008cx}. The modification of the real part of the potential also does not always correspond to an exponentially screened form~\cite{Brambilla:2008cx}; specifically, for interquark distances less than the Debye screening length,  the potential is altered by power-law thermal effects rather than exponential screening. 
Indeed, while color screening is well-established at high temperature in terms of the free energy of a static quark anti-quark pair through lattice calculations~\cite{Bazavov:2018wmo}, its manifestation in the real part of the heavy quark potential is still inconclusive.
In a weakly coupled quark-gluon plasma the imaginary part of the heavy quark potential leads to a thermal quarkonium yield, and in fact is the dominant source of the disappearance of different quarkonium states~\cite{Escobedo:2008sy,Laine:2008cf}. Recent lattice QCD studies of the complex potential at nonzero temperature, assuming a dominant spectral peak, do not find substantial evidence for color screening till distances $\lesssim 1$~fm, contrary to screened real part reported by studies based on Hard-Thermal-Loop inspired ansatz, although all find a large imaginary part \cite{Bala:2021fkm,Bazavov:2023dci,Tang:2023tkm}. Irrespective of the detailed microscopic mechanism, we still expect all the quarkonium states to be dissociated (melted) at sufficiently high temperatures. This is consistent with the behavior of the spatial quarkonium correlation function at high temperatures~\cite{Karsch:2012na,Bazavov:2014cta,Petreczky:2021zmz}.

In-medium quarkonium properties are encoded in the corresponding spectral functions, which are related to the Euclidean time correlation functions calculable on the lattice. There have been many attempts to study in-medium properties of charmonium~\cite{Umeda:2002vr,Datta:2002ck,Karsch:2002wv,Datta:2003ww,Asakawa:2003re,Jakovac:2006sf,Ohno:2011zc,Ding:2012sp,Ding:2017std} and bottomonium~\cite{Jakovac:2006sf,Aarts:2010ek,Aarts:2011sm,Aarts:2012ka,Aarts:2013kaa,Aarts:2014cda,Kim:2014iga,Kim:2018yhk} in lattice QCD. These studies focused on in-medium modifications of ground states of S- and P-wave quarkonia. Previous lattice QCD studies of in-medium quarkonium have mostly used point meson operators, i.e., operators with quark and anti-quark fields located in the same spatial point, which are known to have non-optimal overlap with the quarkonium wave-functions, especially, with the excited states. As a result, these correlators are largely dominated by the vacuum continuum parts of the spectral function, and isolating the contributions from the  in-medium bottomonium modification is very difficult~\cite{Mocsy:2007yj,Petreczky:2010tk,Burnier:2015tda}. 
Recently, the possibility of studying in-medium bottomonium properties using correlators of extended meson operators has been explored within the framework of non-relativistic QCD (NRQCD)~\cite{Larsen:2019bwy,Larsen:2019zqv,Larsen:2020rjk}. It was found that at zero temperature such operators have very good overlap with different S- and P-wave bottomonia~\cite{Davies:1994mp,Meinel:2009rd,Meinel:2010pv,Hammant:2011bt,HPQCD:2011qwj,Daldrop:2011aa} while remaining unaffected by the continuum, and therefore, their correlators are expected to be more sensitive to in-medium bottomonium properties than the ones with point sources.
Furthermore, at finite temperatures these extended operators are anticipated to preserve their utility, as the short-distance physics is largely unaltered by the thermal effects. Accordingly, these calculations at finite temperatures found that the in-medium mass shifts  are consistent with zero, but the thermal bottomonium widths are large \cite{Larsen:2019bwy,Larsen:2019zqv}. This is consistent with the absence of screening in the real part of the potential up to $\lesssim 1$~fm distances, but a sizable imaginary part~\cite{Shi:2021qri, Tang:2024dkz}. In fact, the estimates of the width obtained in Refs.~\cite{Larsen:2019bwy,Larsen:2019zqv} are larger than most phenomenological estimates; see Ref.~\cite{Andronic:2024oxz}. 
However, the above calculations have been performed on lattices with temporal extent $N_{\tau}=12$ and a single lattice spacing for any given temperature values. Therefore, it is very important to extend these calculations to larger temporal extent and having more than one lattice spacing per temperature value. Furthermore, the temperature region below $199$ MeV was not explored. In this paper, we present calculations of bottomonium correlation functions using NRQCD and extended meson operators in the temperature range $T=133-250$ MeV using lattices with temporal extent $N_{\tau}=16-30$ and two lattice spacings $a=0.0493$ and $0.0602$ fm.

The remainder of the paper is organized as follows. In the next section, we discuss the lattice calculations of bottomonium correlators, including our lattice QCD setup.
In \autoref{sec:in-med} we discuss the determination of in-medium bottomonium
masses and width. Finally, \autoref{sec:concl} contains our conclusions. Some technical
details, including the fitting processes to extract vacuum masses and in-medium parameters, are discussed in the appendix.

\section{The lattice NRQCD calculations of bottomonium correlators}

Because of the large bottom ($b$) quark mass, the bottomonium system can be described in terms of NRQCD, an effective field theory where the energy scale associated with the heavy quark mass is
integrated out~\cite{Caswell:1985ui,Thacker:1990bm}. This effective field theory can be formulated on the lattice,
and lattice NRQCD has been successfully used in studying bottomonium properties in  vacuum~\cite{Lepage:1992tx,Davies:1994mp,Meinel:2009rd,Hammant:2011bt,HPQCD:2011qwj,Daldrop:2011aa}, since the large discretization effects associated with the large $b$ quark mass encountered in the relativistic formulation are avoided.
NRQCD has also been used to study bottomonium properties at nonzero temperatures~\cite{Aarts:2010ek,Aarts:2011sm,Aarts:2012ka,Aarts:2013kaa,Aarts:2014cda,Kim:2014iga,Kim:2018yhk,Larsen:2019bwy,Larsen:2019zqv}. In this section, we describe our calculation of bottomonium correlators with extended meson operators at zero temperature and finite temperatures. Extended meson operators offer improved projection onto specific bottomonium states compared to point-like counterparts, making correlators less sensitive to the continuum~\cite{Davies:1994mp,Meinel:2009rd,Meinel:2010pv,Hammant:2011bt,HPQCD:2011qwj,Daldrop:2011aa,Larsen:2019bwy,Larsen:2019zqv}. For this reason, the corresponding
correlators are also more sensitive to the in-medium modification of bottomonium
properties.

In \autoref{subsec:setup} we discuss our lattice setup, including the 
extended bottomonium operators used in our study. Our results at zero temperature,  including the bottomonium mass spectrum, are discussed in~\autoref{subsec:vacuum}. The temperature dependence of the bottomonium correlators is discussed in~\autoref{subsec:in-medium}.

\subsection{The calculation details \label{subsec:setup}}
The lattice calculations in this work are carried out on background gauge fields including (2+1)-flavor dynamical sea quarks, which are implemented using the highly improved staggered quarks and the tree-level Symanzik gauge (HISQ/tree) action. The strange quark mass $\ms$ is fixed to its physical value, with light quark masses equal to $\ms/20$, corresponding to the Goldstone pion mass about 160 MeV, which is near the physical point. In this work, the spatial extents of the lattices are set to $N_{\sigma}=64$ and the temperature is varied by varying the temporal extent $N_{\tau}$ from $30$ to 16 at a fixed lattice spacing $a$. We perform calculations at values of the lattice
gauge coupling $\beta=10/g_0^2=7.596$ and 7.373, corresponding to lattice spacings $a=0.0493$ and $a=0.0602$ fm, respectively. 
The physical value of the lattice spacing is fixed using the parametrization of the $r_1$ scale from Ref. \cite{Bazavov:2017dsy} and the value $r_1=0.3106(17)$ fm from Ref.~\cite{MILC:2010hzw}. 
The gauge configurations were generated using the software suite \texttt{SIMULATeQCD}~\cite{Altenkort:2021cvg}.

With these lattice spacings and $N_{\tau}$ values, we cover the temperature range from 133 MeV to 250 MeV. We also perform calculations on $48^3 \times 64$ and $64^4$ lattices for $a=0.0602$ and $a=0.0493$ fm, respectively, to provide a zero-temperature baseline, using gauge configurations
generated by the HotQCD collaboration~\cite{HotQCD:2014kol}.
\begin{table}[ht]
\centering
\resizebox{0.9\textwidth}{!}{
\begin{tabular}{*{8}{c}}
\hline
\hline
\multirow{2}{*}{$\beta$}                          &
\multirow{2}{*}{$a$ [fm]}                         &
\multirow{2}{*}{$N_{\sigma}^3 \times N_{\tau}$}   & 
\multirow{2}{*}{$T$ [MeV]}                        & 
\multirow{2}{*}{$u_0$}                            &
\multirow{2}{*}{$aM_b$}                           &
\multicolumn{2}{c}{$\#$ conf.} 
\\  
\cline{7-8}
&&&&&& Gaussian src. & Wave-opt. src.
\\
\hline \hline
\vspace{-3ex}
\multirow{11}{*}{7.596}  & \multirow{11}{*}{0.0493} & & & \multirow{11}{*}{0.89517} & \multirow{11}{*}{0.957} &
\\
 & & $64^3\times$16 & 249.8 & & & 1584 & 4412
\\
 & & $64^3\times$18 & 222.2 & & & 1583 & 3469
\\
 & & $64^3\times$20 & 200.0 & & & 1596 & 3000
\\
 & & $64^3\times$22 & 181.8 & & & 1596 & 3924
\\
 & & $64^3\times$24 & 166.6 & & & 1536 & 3235
\\
 & & $64^3\times$26 & 153.8 & & & 1536 & 2391
\\
 & & $64^3\times$28 & 142.8 & & & 1524 & 1997
\\
 & & $64^3\times$30 & 133.3 & & & 1608 & 1608
\\
 & & $64^3\times$64 & $\simeq 0$ & & & 1152 & 1152
\\
\hline
\vspace{-3ex}
\multirow{6}{*}{7.373}  & \multirow{6}{*}{0.0602} & & & \multirow{6}{*}{0.89035} & \multirow{6}{*}{1.22} &
\\
 & & $64^3\times$16 & 204.9 & & & -- & 3770
\\
 & & $64^3\times$18 & 182.1 & & & -- & 2800
\\
 & & $64^3\times$20 & 163.4 & & & -- & 3660
\\
 & & $64^3\times$24 & 136.6 & & & -- & 2268
\\
 & & $48^3\times$64 & $\simeq 0$ & & & -- & 564
\\
\hline
\end{tabular}}
\caption{Summary of lattice parameters; i.e., the lattice gauge coupling $\beta$, the lattice spacing $a$, the lattice size $N_{\sigma}^3 \times N_{\tau}$, the temperature $T$, the tadpole parameters $u_0$, the tuned bare bottom quark mass $aM_b$ and the number of gauge configurations ($\#$conf.) measured with Gaussian and wave-function-optimized sources, respectively.}
\label{tab:setup}
\end{table}

For the bottom quark on the lattice, the tree-level tadpole-improved NRQCD action including spin-dependent $v^6$ corrections is employed, i.e. exactly the same setup as in Refs.~\cite{Meinel:2010pv,Larsen:2019bwy,Larsen:2019zqv}. Here $v$ is the heavy-quark velocity inside quarkonium. The tadpole improvement is achieved by dividing the link variables by the tadpole parameter $u_0$. In our calculations, we take $u_0$ to be the fourth root of the averaged plaquette. The stability or Lepage parameter $n$, which controls the time discretization steps in the evolution of heavy-quark propagator, is set to $n=16$. This value is larger than the ones used in Refs. \cite{Larsen:2019bwy,Larsen:2019zqv}. Given the relatively small lattice spacings we found it is safer to use large values of $n$.

We utilize two types of extended meson operators proposed in Refs.~\cite{Larsen:2019bwy,Larsen:2019zqv}: one employs Gaussian-shaped smearing for ground S- and P-wave bottomonium states, while the other is optimized using wave-functions obtained by solving the discretized \Schrodinger equation with a Cornell potential at zero temperature, as specified in Refs.~\cite{Meinel:2010pv,Larsen:2019zqv}, for different quarkonium states.
More specifically, we use the Gaussian-smeared extended meson operator given by $O(\mathbf{x},\tau)=\tilde{\bar{q}}(\mathbf{x},\tau)\Gamma\tilde{q}(\mathbf{x},\tau)$, where $\Gamma$ denotes the bottomonium interpolators described in Ref.~\cite{Larsen:2019bwy} and $\tilde{q}$ and $\tilde{\bar q}$ stand for smeared quark and anti-quark fields. The smeared quark field is obtained from the original quark field as
$\tilde{q}=W q$~\cite{Larsen:2019bwy},
where $W=\left[1+ \sigma^2 \Delta^{(2)} / (4N)\right]^N$.
For large enough $N$, this produces a smeared quark field with Gaussian profile having a width $\sigma$. We adopt the same smearing parameters $\sigma$ and $N$ to construct $W$ as those specified in Ref.~\cite{Larsen:2019bwy}. 
To construct the  wave-function-optimized meson operator,
we first consider meson operators of the form $O_{\alpha}(\mathbf{x}, \tau)=\sum_{\mathbf{r}} \Psi_{\alpha}(\mathbf{r}) \bar{q}(\mathbf{x}+\mathbf{r}, \tau) \Gamma q(\mathbf{x}, \tau)$~\cite{Larsen:2019zqv}, where $\alpha$ represents different excited states for a specific bottomonium channel, ranging from 1S(P) to 3S(P) states in our study. The wave-function $\Psi_{\alpha}(\mathbf{r})$ is solved from the discretized three-dimensional \Schrodinger equation, where an $\mathcal{O}(a^4)$-improved discretized Laplacian is used with the same spacing as the gauge backgrounds, and the potential takes a Cornell form with the same parameterization as in Refs.~\cite{Meinel:2010pv,Larsen:2019zqv}.
Then, using the correlators of the above meson operators, we perform a variational analysis to effectively eliminate nonzero off-diagonal correlators $C_{\alpha\beta}(\tau) = \Sigma_{\mathbf{x}} \langle O_{\alpha}(\mathbf{x}, \tau) O^{\dagger}_{\beta}(\mathbf{0}, 0) \rangle$ ($\alpha\neq\beta$). Through solving for the rotation matrix $\Omega_{i\beta}$ from the generalized eigenvalue problem~\cite{Larsen:2019zqv,Blossier:2009kd,Orginos:2015tha,Nochi:2016wqg}, formulated as $C_{\alpha\beta}(\tau) \Omega_{i \beta}=\lambda_{i}\left(\tau, \tau_{0}\right) C_{\alpha\beta}(\tau_0) \Omega_{i\beta}$, we obtain the diagonalized correlator $C_i(\tau)=\sum_{\mathbf{x}}\langle \Omega_{i\alpha} O_{\alpha}(\mathbf{x},\tau) O^{\dagger}_{\beta}(\mathbf{0},0)\Omega^{\dagger}_{\beta i}\rangle$. Here $\alpha$, $\beta$ and $i$ represent different excited states, and the summations are over $\alpha$ and $\beta$. The rotation matrix $\Omega_{i\beta}$ is computed at zero temperature and uniformly applied across all temperatures. During this computation, the $\tau$ is fixed at about 0.5 fm, while $\tau_0$ is set to be either  0 or $4 a$. 
We use two different $\tau_0$ to explore the influence of $\tau_0$ choice on our results.
We find that the choice of $\tau_0$ has no effect on the bottomonium correlation functions  
except for $\tau \simeq 1/T$ for higher temperatures.

For both types of extended operators mentioned above, bottomonium correlators are measured using 24 sources distributed across different locations for each gauge configuration to improve the signal. 

The bottom quark mass, $M_b$, appearing in the NRQCD action, is fixed by demanding that the kinetic mass of $\eta_b$, defined as $E_0(M_b, \mathbf{p}) =\sqrt{\mathbf{p}^{2}+M_{\mathrm{kin}}^{2}(M_b)}+\text{const}$, agrees with the value of the $\eta_b$ mass from the Particle Data Group. We use the values of $M_b$ determined in~\cite{Larsen:2019bwy}. For $\beta=7.596$, we performed an independent analysis of the kinetic mass and found a value of $aM_b=0.955(17)$, which is consistent with the previous result $aM_b=0.957(9)$~\cite{Larsen:2019bwy}.
The parameters of the lattice NRQCD calculations are summarized in \autoref{tab:setup}.

\subsection{Results in vacuum \label{subsec:vacuum}}

\begin{figure}[!htbp]
\centering
    \includegraphics[width=0.45\textwidth]{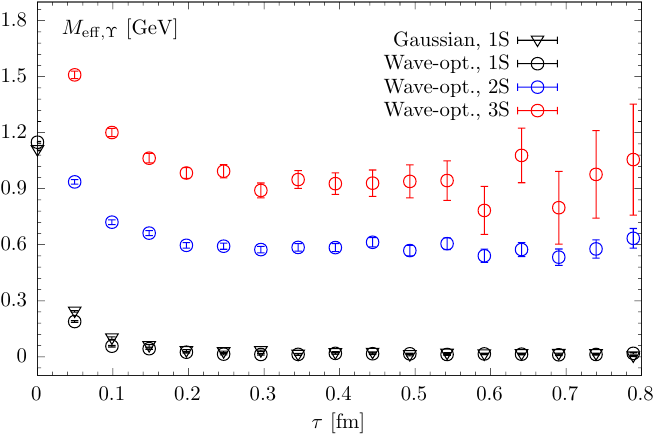}
    \qquad
    \includegraphics[width=0.45\textwidth]{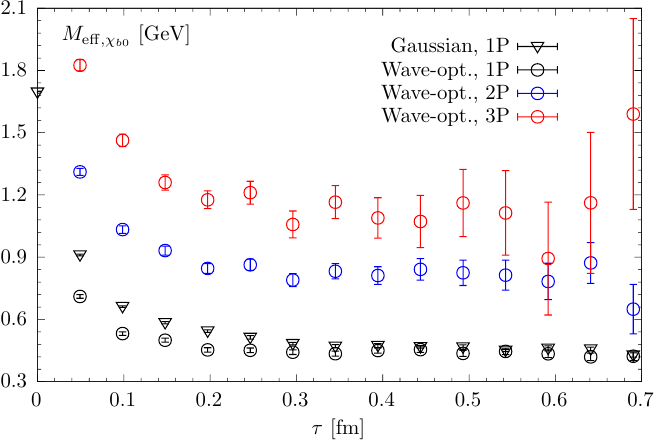}
  	\caption{The effective masses $\meff$ at zero temperature for $\Upsilon$ (left) and $\chi_{b0}$ (right) states, from Gaussian-smeared extended sources (triangle) and wave-function-optimized sources (circle) respectively. The vertical scale is calibrated by subtracting the spin-averaged mass of 1S bottomonia, $\bar{M}_{\rm 1S}=(M_{\eta_b}+3M_{\Upsilon})/4$, at $T=0$.}
	\label{fig:meff-zero-temperature}
\end{figure}

Our numerical results on the bottomonium correlation functions, $C(\tau)$, at zero temperature are shown in \autoref{fig:meff-zero-temperature}, in terms of the effective masses defined as $aM_{\text{eff}}(\tau)=\log[C(\tau)/C(\tau+a)]$. 
Hereafter, the scale for the effective mass plot is typically calibrated by subtracting the spin-averaged mass of 1S bottomonia at $T=0$, unless otherwise specified. We show results for both Gaussian and wave-function-optimized operators at $\beta=7.596$. Our results at $\beta=7.373$ look similar. Both Gaussian and wave-function-optimized operators are effective in suppressing the excited state contributions for 1S and 1P bottomonium states. There are some differences in the effective masses corresponding to the two types of meson operators at very small $\tau$, but they approach the same plateau. For the 1S state, we see that the plateau is reached already at $\tau\simeq 0.25$ fm. The wave-function-optimized sources for 2S, 3S, 2P, and 3P also show clear plateaus for $\tau \simeq 0.3$ fm. Since the plateaus in the effective masses are reached at time separations smaller than the inverse temperature, the correlators of extended operators are suitable to constrain the in-medium bottomonium properties.

\begin{figure}[!htbp]
  \centering
    \includegraphics[width=0.7\textwidth]{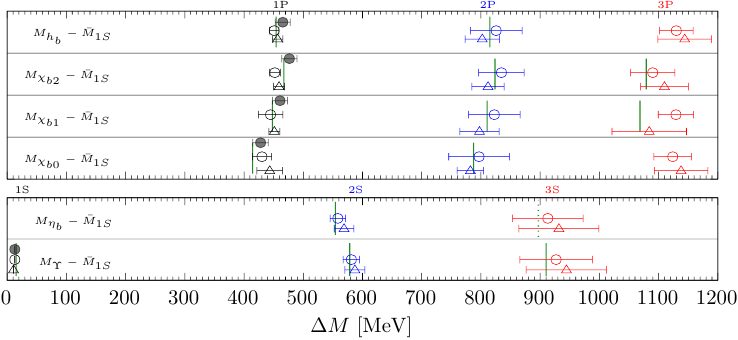}
  	\caption{Comparisons of mass differences $\Delta M$ for different bottomonium states with respect to the 1S spin-averaged mass $\bar{M}_{1S}$ from lattice NRQCD calculations (points) with those from PDG~\cite{ParticleDataGroup:2024cfk} (green solid lines). Green dashed line is the lattice-predicted value from Ref.~\cite{Larsen:2019zqv}. Open points are obtained from analyses on 1S/1P (black), 2S/2P (blue) and 3S/3P (red) wave-function-optimized sources (circle for $a=0.0493$ fm and triangle for $0.0602$ fm), while filled ones from Gaussian sources.}
	\label{fig:mass-difference}
\end{figure}

From the large $\tau$ behavior, we extract the energy levels of the different bottomonium states. We fit the correlators
with a one-state fit in the interval $[\tau_{\rm min}:\tau_{\rm max}]$. The value of $\tau_{\rm min}$ is chosen such that the excited-state contributions are significantly smaller than the statistical errors, while $\tau_{\rm max}$ is chosen such that the signal-to-noise ratio is good. The details of the fit procedure are given in Appendix~\ref{appendix:vacuum-mass-extraction}. In NRQCD calculations, the energy levels
cannot be directly compared to the bottomonium masses observed in experiment. However, the difference in the energy
levels of different bottomonium states correspond to the difference of the experimentally measured bottomonium masses.
Therefore, in \autoref{fig:mass-difference} we show the mass differences
between different states and the 1S spin-averaged mass $\bar{M}_{\rm 1S}=(M_{\eta_b}+3M_{\Upsilon})/4$, calculated in NRQCD and compared to the 
experimental results from PDG. The reason for calculating the mass differences with
respect to the spin-averaged 1S mass is that the effects of the hyperfine splitting do not enter.
Reproducing the hyperfine splitting of 1S bottomonium in lattice NRQCD is difficult because it is very sensitive to the short-distance physics. As one can see from \autoref{fig:mass-difference}, our NRQCD calculations can reproduce the experimental results on the differences
within errors.

\subsection{Temperature dependence of bottomonium correlators \label{subsec:in-medium}}

\begin{figure}[htbp]
\centering
    \includegraphics[width=0.45\textwidth]{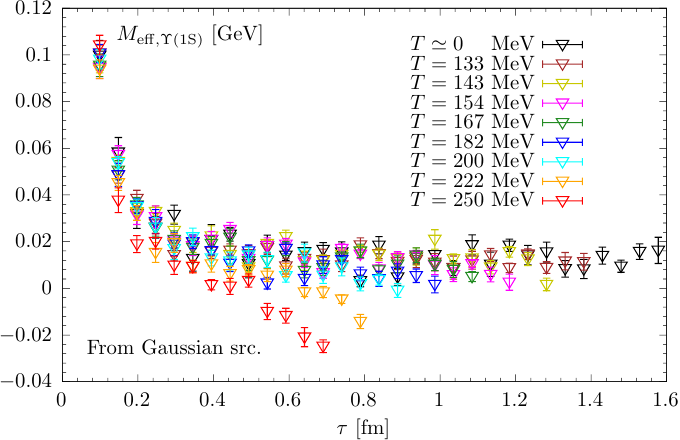}
    \qquad
    \includegraphics[width=0.45\textwidth]{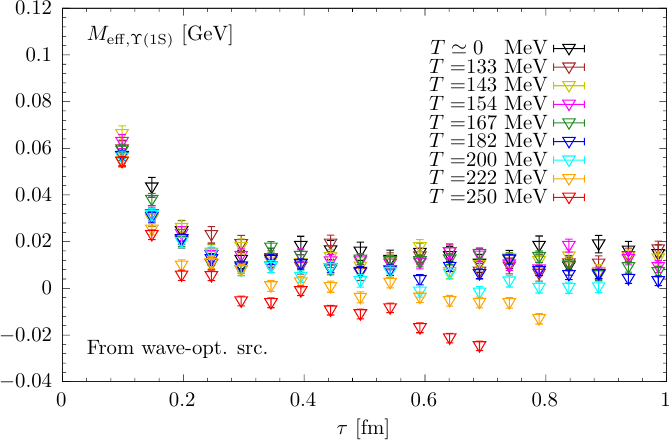}
    \includegraphics[width=0.45\textwidth]{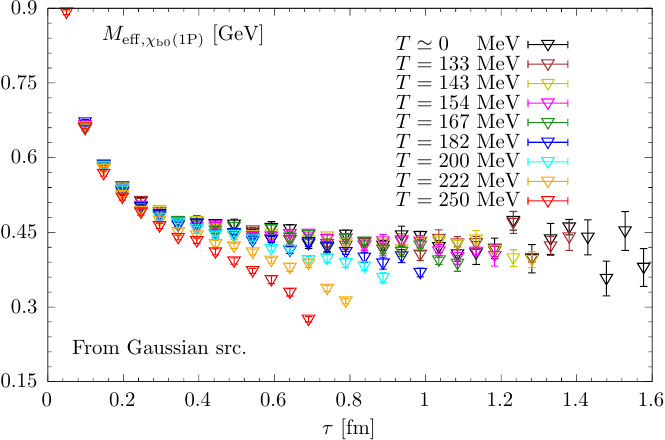}
    \qquad
    \includegraphics[width=0.45\textwidth]{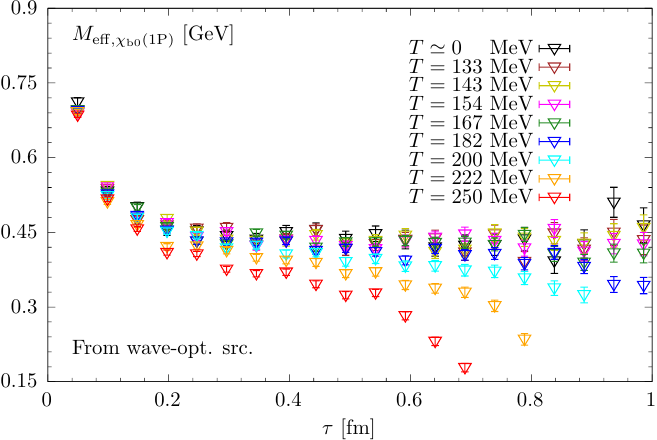}
    \caption{Effective masses for the $\Upsilon$(1S) (top) and the $\chi_{b0}$(1P) (bottom) states. The left panel shows the results from Gaussian sources, while the right panel shows the results from the wave-function-optimized sources. The vertical scale is calibrated by subtracting the spin-averaged mass of 1S bottomonia at $T=0$.}
    \label{fig:meff-T-1S1P}
\end{figure}

The temperature dependence of bottomonium correlation functions can be conveniently studied in terms
of the effective masses. In \autoref{fig:meff-T-1S1P} we show the effective masses from
Gaussian and wave-function-optimized sources for the $\Upsilon$(1S) and $\chi_{b0}$(1P)
states. At small $\tau$, we see a rather weak temperature dependence of the effective masses
throughout the entire temperature region. The temperature effects gradually increase with increasing
temperatures and it is larger for the $\chi_{b0}$(1P) state than for the $\Upsilon$(1S).
We see that the effective masses seem to approach a plateau for the lowest five temperatures
for the $\Upsilon$(1S) and for the lowest four temperatures for the $\chi_{b0}$(1P).
This suggests that the $\Upsilon$(1S) and $\chi_{b0}$(1P) can exist as bound states
above the crossover temperature with about the same masses as in vacuum. More precisely, the $\Upsilon$(1S) exists as a well-defined state up to a temperature of about $182$ MeV with nearly the same mass as in the vacuum, while the $\chi_{b0}$(1P) can exist up to $T=167$ MeV with the vacuum mass. At higher temperature, we see a clear decrease in the effective masses, possibly signaling in-medium modifications of the corresponding bottomonium states.
Qualitatively, the temperature dependence of the effective masses obtained using Gaussian 
and wave-function-optimized sources is similar, but there are quantitative differences
that have to be understood in order to determine in-medium bottomonium properties.
This will be discussed in the next section. 

\begin{figure}[htbp]
\centering
    \includegraphics[width=0.45\textwidth]{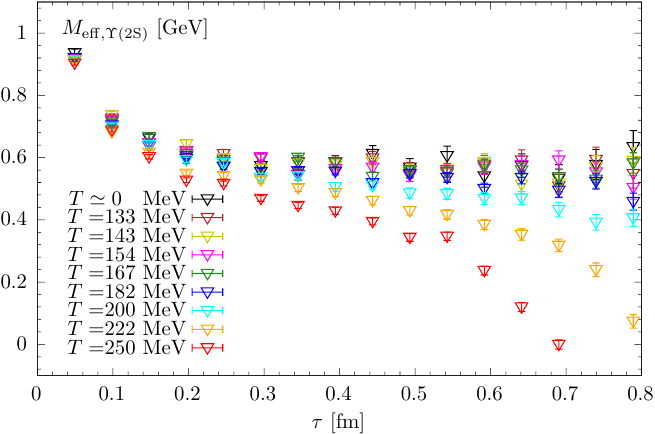}
    \qquad
    \includegraphics[width=0.46\textwidth]{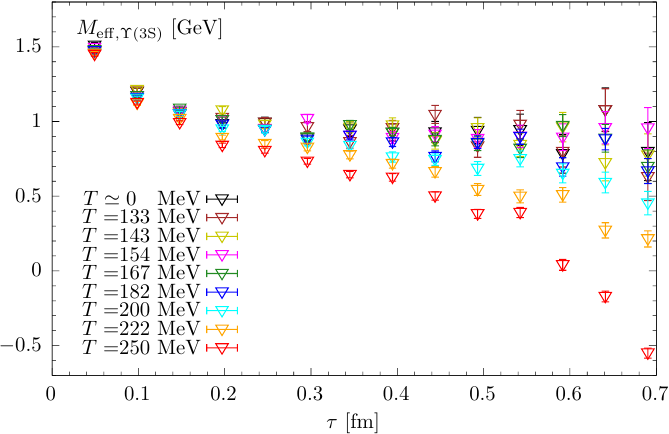}
    \includegraphics[width=0.45\textwidth]{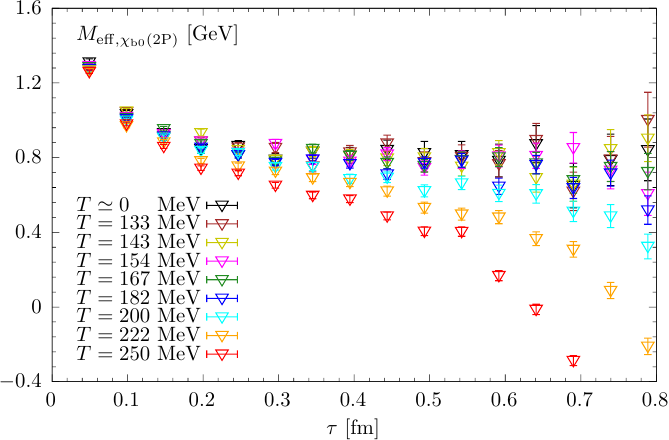}
    \qquad
    \includegraphics[width=0.45\textwidth]{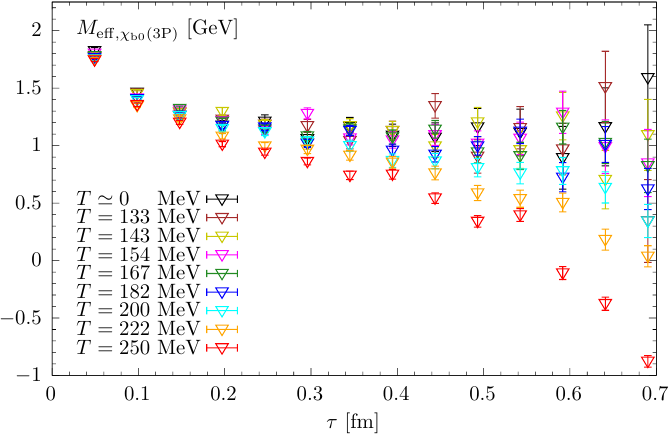}
    \label{fig:meff-finite-temperature-wave}
  	\caption{Temperature dependence of the effective masses for the $\Upsilon$(2S), $\Upsilon$(3S), $\chi_{b0}$(2P) and $\chi_{b0}$(3P) states from measurements using wave-function-optimized sources. The vertical scale is calibrated by subtracting the spin-averaged mass of 1S bottomonia at $T=0$.}
	\label{fig:meff-T_2S2P}
\end{figure}

In \autoref{fig:meff-T_2S2P} we show the effective masses from wave-function-optimized
sources for the $\Upsilon$(2S), $\Upsilon$(3S), $\chi_{b0}$(2P) and $\chi_{b0}$(3P).
Again we see that at small $\tau$ the temperature dependence is weak and increases
with increasing $\tau$. At low temperature, namely $T<167$ MeV, we see indications of
a plateau in the effective masses, indicating that these states can exist at least up to this 
temperature. Above that temperature we again clearly see the temperature dependence of the corresponding
effective masses.
The temperature dependence of the effective masses for the
$\Upsilon$(3S) is larger than the temperature dependence of the effective masses
for the $\Upsilon$(2S). This is expected since the $\Upsilon$(3S) is larger in size
than the $\Upsilon$(2S), and therefore will be more affected by the medium.
In general we find that the effective masses 
corresponding to higher excited states, which are larger in size, are more affected by
the medium in line with common expectations.

\section{Bottomonium properties at nonzero temperature \label{sec:in-med}}

Deeper insights into the in-medium properties of bottomonia are encoded in the spectral function $\rho(\omega,T)$, which is related to the correlators by a Laplace transform
\begin{align}
	C(\tau, T)=\int_{-\infty}^{+\infty} \mathrm{d} \omega ~ \rho(\omega, T) e^{-\tau \omega}\,.
	\label{eq:spectra-function-definition}
\end{align}
For the bottomonium system studied here, the spectral function is generally expected to contain several states below the open-bottom threshold, with numerous higher states forming a continuum at large frequencies. Thus, $\rho(\omega,T)$ can be decomposed into two parts: $\rho(\omega,T)=\rho_{\mathrm{med}}(\omega,T)+\rho_{\mathrm{cont}}(\omega)$. The continuum part, $\rho_{\mathrm{cont}}(\omega)$, is considered to be temperature independent, aligning with the above observations that the effective masses exhibit almost no temperature dependence at $\tau\lesssim$0.16 fm. On the other hand, the extended sources used in our work are designed to selectively enhance the overlap with particular states while suppressing contributions from undesired ones~\cite{Larsen:2019zqv,Larsen:2019bwy,Wetzorke:2001dk,Stickan:2003gh}. As a result, $\rho_{\mathrm{med}}(\omega,T)$ in vacuum can be parametrized as $\rho_{\mathrm{med}}(\omega,T=0)=A\delta(\omega-M)$, where $M$ represents the mass of a particular state targeted for projection. Combining these considerations, the continuum part can be isolated at zero temperature by $\rho_{\mathrm{cont}}(\omega)=\rho(\omega,T=0)-A\delta(\omega-M)$, leading to 
\begin{align}
	C_{\mathrm{cont}}(\tau)=C(\tau, T=0)-Ae^{-M\tau}\,.
	\label{eq:correlator-continuum}
\end{align}
Here \autoref{eq:spectra-function-definition} is employed and $C_{\mathrm{cont}}(\tau)$ is defined as the Laplace transform of $\rho_{\mathrm{cont}}(\omega)$. The parameters $A$ and $M$ are extracted by fitting relevant correlators in vacuum, implemented as outlined in Appendix~\ref{appendix:vacuum-mass-extraction}. Subsequently, a continuum-subtracted correlator can be defined as in Refs.~\cite{Larsen:2019zqv,Larsen:2019bwy}, 
\begin{align}
	C_{\mathrm{sub}}(\tau,T) = C(\tau, T) - C_{\mathrm{cont}}(\tau)\,.
    \label{eq:continuum-subtracted-correlator}
\end{align}
This preserves the contribution from $\rho_{\mathrm{med}}(\omega,T)$ to the greatest extent, thereby reflecting the in-medium effects in their purity. The continuum-subtracted effective mass is then defined as follows for further extraction of in-medium properties:
\begin{align}
	M_{\mathrm{eff}}^{\mathrm{sub}}(\tau, T) =\log \left[\frac{C_{\mathrm{sub}}\left(\tau, T\right)}{C_{\mathrm{sub}}\left(\tau+a, T\right)}\right]\,.
	\label{eq:continuum-subtracted-effective-mass}
\end{align}

\subsection{Continuum-subtracted effective masses and in-medium bottomonium properties from simple ans\"atze \label{sec:msub-and-fits}}

\begin{figure}[htbp]
\centering
  	\includegraphics[width=0.45\textwidth]{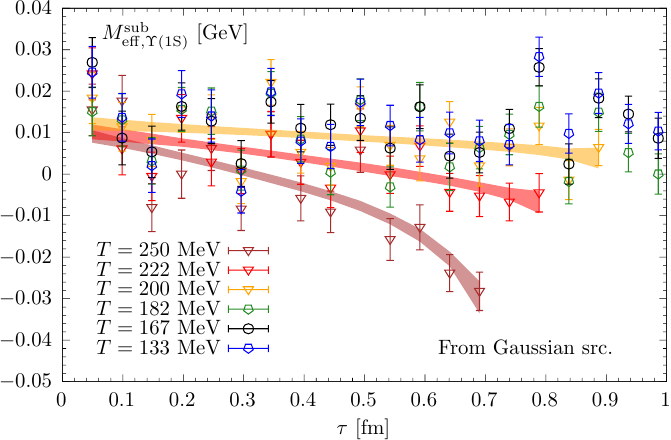}
    \qquad
    \includegraphics[width=0.44\textwidth]{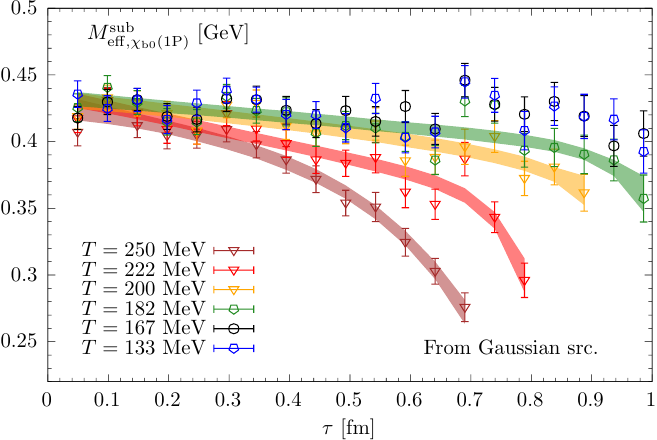}
    \includegraphics[width=0.45\textwidth]{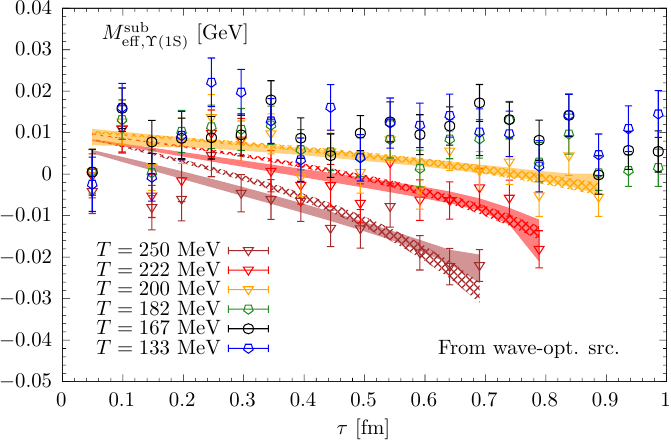}
    \qquad
	\includegraphics[width=0.44\textwidth]{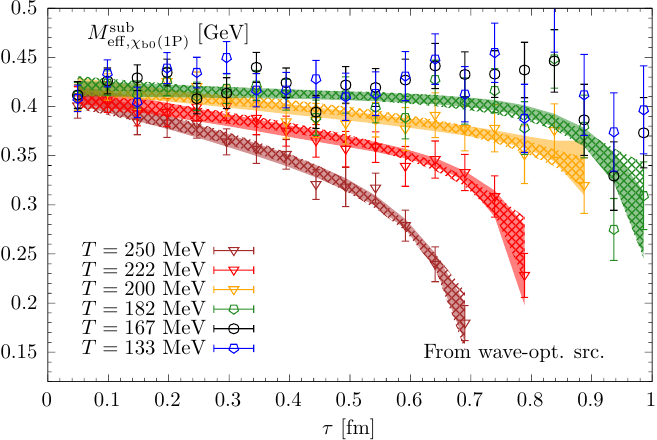}
    \caption{The continuum-subtracted effective masses at different temperatures for the $\Upsilon$(1S) (left) and $\chi_{b0}$(1P) (right) states. The top panel shows the results from Gaussian sources, while the bottom panel shows the results from the wave-function-optimized sources. The solid bands are reconstructed from fits using the cut Lorentzian ans\"atze including the delta term with $cut=4\Gamma_0$; the bands with pattern are from fits using the smooth-cut Lorentzian ans\"atze. The vertical scale is calibrated by subtracting the spin-averaged mass of 1S bottomonia at $T=0$.}
    \label{fig:subtracted-meff-1S1P}
\end{figure}

\begin{figure}[!htbp]
\centering
	\includegraphics[width=0.45\textwidth]{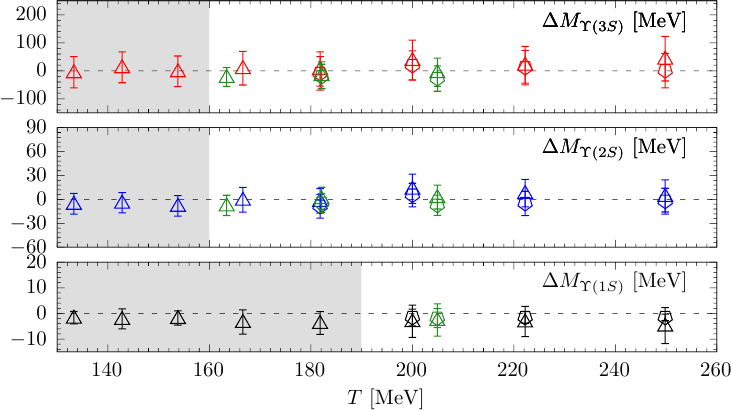}
    \qquad
	\includegraphics[width=0.45\textwidth]{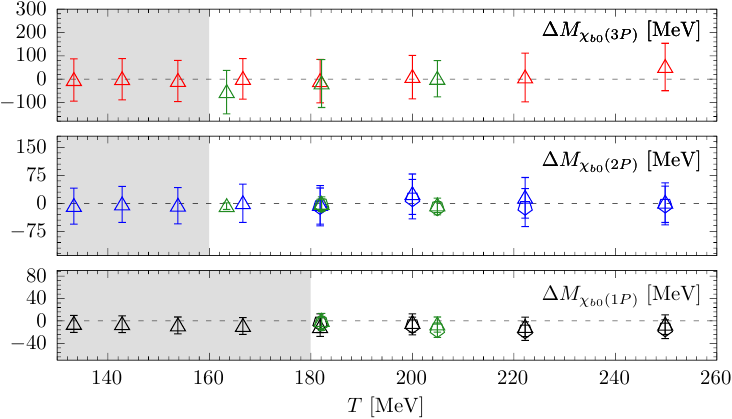}
	\caption{Temperature dependence of in-medium mass shifts for the $\Upsilon$ and $\chi_{b0}$ states, given by cut Lorentzian fits with $cut=4\Gamma_0$ (triangle) and T-matrix motivated smooth-cut Lorentzian fits (pentagon). Green points are results with spacing equal to 0.0602 fm, while others correspond to 0.0493 fm. The results within the gray-shaded areas are from constant fits on corresponding continuum-subtracted effective masses.}
	\label{fig:inmedium-mass-shift}
\end{figure}

The continuum-subtracted effective masses for the $\Upsilon$(1S) and $\chi_{b0}$(1P) states from Gaussian and wave-function-optimized sources are shown in \autoref{fig:subtracted-meff-1S1P}. The continuum-subtracted effective masses show a nearly constant behavior at the lowest two temperatures.
For $\Upsilon$(1S) state the continuum-subtracted effective masses are compatible with a constant behavior even at $T=182$ MeV. Therefore, we can extract
the masses of the corresponding bottomonium states at these temperatures by fitting the continuum-subtracted effective masses with constants, or, equivalently, fitting the continuum-subtracted correlation functions with a single exponential. This is discussed in Appendix \ref{appendix:inmedium-parameters-extraction} in more detail. 
In \autoref{fig:inmedium-mass-shift} we show the results for the in-medium masses, $M_{\rm med}$, in terms of the mass shifts
\begin{equation}
    \Delta M(T)=M_{\rm med}(T)-M(T=0).
\end{equation}
As one can see from the figure, $\Delta M$ is compatible with zero within errors, i.e., we do
not see any in-medium modification of the $\Upsilon$(1S) bottomonium mass
up to $T=182$ MeV and any in-medium modification of the $\chi_{b0}$(1P) bottomonium mass up
to $T=167$ MeV based on these fits.

At higher temperatures we see that the continuum-subtracted effective masses decrease linearly in $\tau$ if $\tau$ is not too large. For $\tau$
values close to $1/T$ the continuum-subtracted effective masses show a sudden drop in many cases. It was suggested that the linear decrease of the continuum-subtracted effective masses is related to the width of the dominant peak in $\rho_{\rm med}(\omega,T)$~\cite{Larsen:2019bwy,Larsen:2019zqv}. If 
$\rho_{\rm med}(\omega,T)$ would consist of a single Gaussian peak, 
\begin{align}
	\rho_{\mathrm{med}}(\omega, T) =
	 A_{\mathrm{med}}(T) \exp \left(-\frac{\left[\omega-M_{\mathrm{med}}(T)\right]^{2}}{2 \Gamma_{\mathrm{G}}^{2}(T)}\right)\,,
	\label{eq:Gaussian}	
\end{align}
then the slope of the continuum-subtracted effective masses would be given by $\Gamma_G^2$.
Note, however, that a Gaussian form of the spectral function is not the only one that could 
result in nearly linear decrease of the continuum-subtracted effective masses, as discussed below. In NRQCD, there is an additional contribution to the spectral function well below the dominant peak, which corresponds to the forward-propagating heavy $q\bar q$ pair interacting with the background of backward-propagating light degrees of freedom from the heat bath~\cite{Bala:2021fkm}. This contribution will depend on the choice of the meson operator, as different meson operators may have different overlap with the multi-particle states~\cite{Bala:2021fkm}. From \autoref{fig:subtracted-meff-1S1P} we see that the rapid drop-off in the continuum-subtracted effective masses is different for Gaussian and wave-function-optimized sources, as expected. The slope of the linear decrease appears to be the same for both type of sources. Furthermore, the slope of the continuum-subtracted effective masses is larger for the $\chi_{b0}$(1P) state than for the $\Upsilon$(1S) state, possibly suggesting that the in-medium width for the $\chi_{b0}$(1P) state is larger than the one for the $\Upsilon$(1S) state. The differences between the continuum-subtracted effective masses computed using Gaussian and wave-function-optimized sources arise also due to the additional contributions from the regions of the respective spectral functions below the dominant peaks. As discussed below, these differences in the continuum-subtracted effective masses do not lead to differences in the in-medium bottomonium masses and widths. In addition, the behavior of the continuum-subtracted effective masses is influenced by the detailed choice of
the wave-function-optimized sources, e.g. by the choice of $\tau_0$ in the generalized eigenvalue problem, without affecting the in-medium bottomonium parameters.

\begin{figure}[htbp]
\centering
    \includegraphics[width=0.45\textwidth]{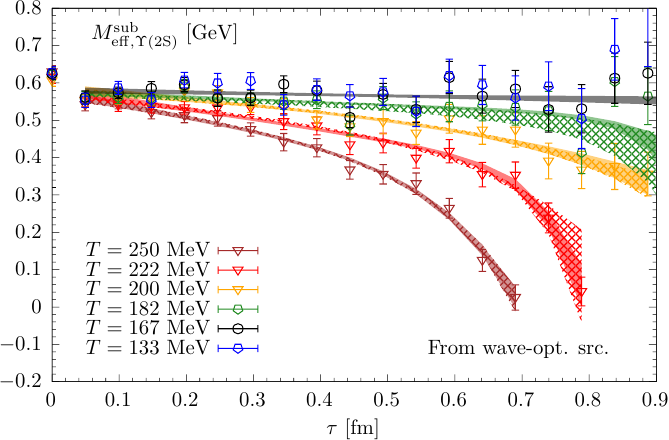}
    \qquad
    \includegraphics[width=0.45\textwidth]{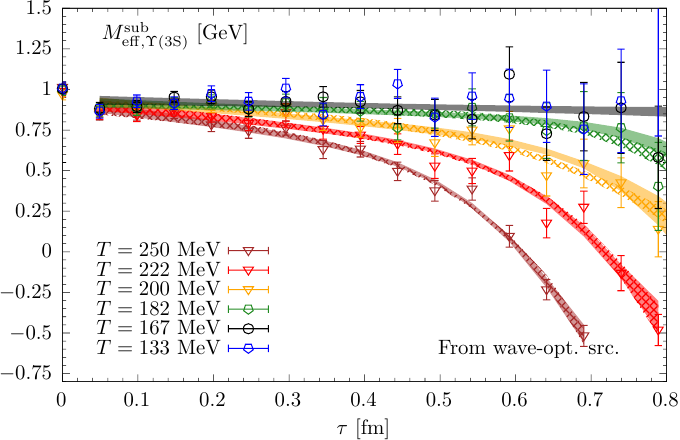}
  	\includegraphics[width=0.45\textwidth]{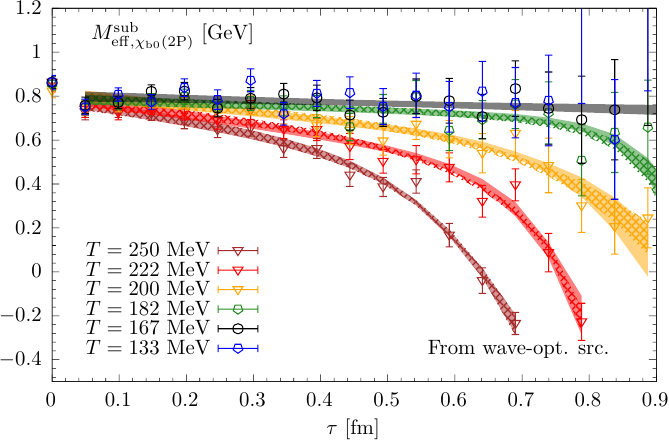}
    \qquad
  	\includegraphics[width=0.45\textwidth]{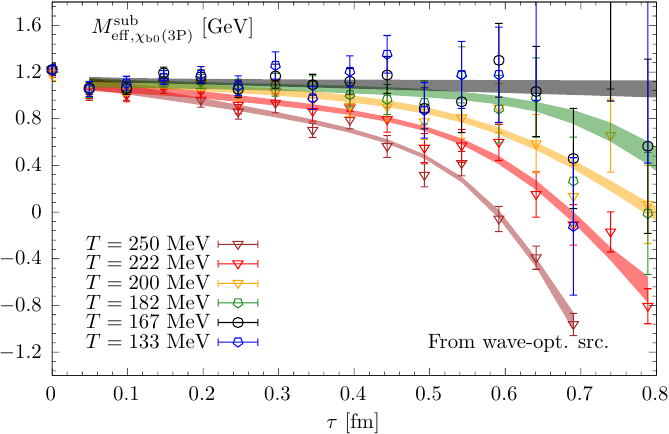}
    \caption{The continuum-subtracted effective masses at different temperatures for excited states of $\Upsilon$ (top) and $\chi_{b0}$ (bottom). The solid bands are reconstructed from fits using the cut Lorentzian ans\"atze including (for $T>180$ MeV) or excluding (for $T<180$ MeV) the delta term, with $cut=4\Gamma_0$; the bands with pattern are from fits using the smooth-cut Lorentzian ans\"atze. The vertical scale is calibrated by subtracting the spin-averaged mass of 1S bottomonia at $T=0$.}
	\label{fig:subtracted-meff-2S3S2P3P}
\end{figure}

In \autoref{fig:subtracted-meff-2S3S2P3P} we show the continuum-subtracted effective masses for the $\Upsilon$(2S), $\Upsilon$(3S), $\chi_{b0}$(2P) and $\chi_{b0}$(3P) states. The continuum-subtracted effective masses are constant within errors at low temperatures and again show a linear decrease at higher temperatures for moderate values of $\tau$. For $\tau$ around $1/T$ the continuum-subtracted effective masses show a rapid drop-off. Note that the slope of the continuum-subtracted effective masses for $\Upsilon$(3S) state is larger than that for the $\Upsilon$(2S) state, and similarly the slope for the $\chi_{b0}$(3P) state is larger than that for the $\chi_{b0}$(2P) state. We again fit the effective masses with a constant, or, equivalently, perform single-exponential fits on the correlators for temperatures below $167$ MeV. The corresponding results for the in-medium bottomonium masses are shown in \autoref{fig:inmedium-mass-shift} in terms of $\Delta M(T)$. We do not see statistically significant mass shifts for the $\Upsilon$(2S), $\Upsilon$(3S), $\chi_{b0}$(2P), and $\chi_{b0}$(3P) states at these temperatures.
To summarize, the continuum-subtracted effective masses indicate that different bottomonium states may exist above the chiral crossover temperature but may acquire a thermal width.
The continuum-subtracted effective masses show a linear decrease at moderate $\tau$ values that can be possibly related to the width of the bottomonium states. The slope of this linear decrease of the continuum-subtracted effective masses with $\tau$ becomes larger with increasing temperature and is also larger for higher-lying bottomonium states. This may indicate that higher-lying bottomonium states, which are larger in size, are more affected by the medium, and thus have larger thermal width, and the thermal width increases with the temperature. The slope of the linear decrease of the continuum-subtracted effective masses is independent of the choice of the meson operators (Gaussian or wave-function-optimized), and therefore likely corresponds
to a physical effect, e.g., thermal broadening of the peak. We also observe a sudden drop in the continuum-subtracted effective masses around $\tau \sim 1/T$, which is not related to the in-medium modification of bottomonium but to the contribution to the spectral function well below the dominant peak.

\begin{figure}[htbp]
\centering
    \includegraphics[width=0.45\textwidth]{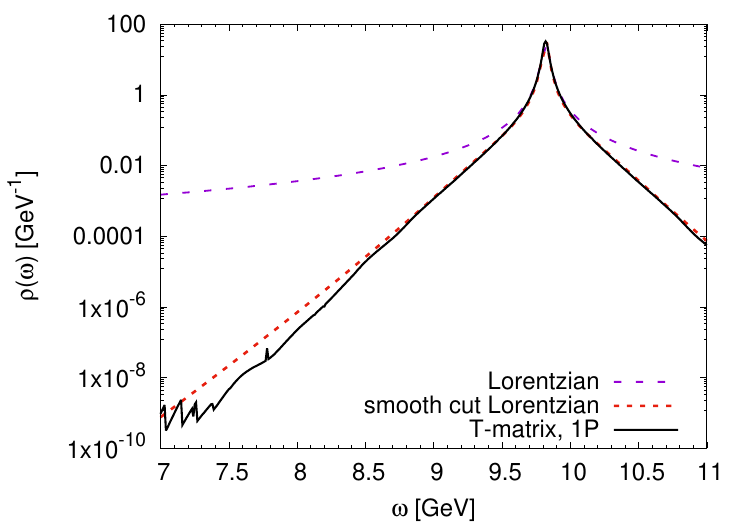}
    \caption{The spectral function of the $\chi_b$(1P) state at $T=182$ MeV calculated in the T-matrix approach, compared with a simple cut Lorentzian parametrization of the spectral function as well as the smooth-cut Lorentzian parametrization.}
\label{fig:spf-example}
\end{figure}

To obtain information on the in-medium properties of different bottomonium states, we need an ans\"atze for $\rho_{\rm med}(\omega,T)$, and, in particular, a parametrization of the dominant peak. It is expected that spectral functions 
at nonzero temperature have a quasi-particle peak described by a Lorentzian form in the vicinity of the peak. This is confirmed by explicit calculations of the bottomonium spectral functions in the T-matrix approach~\cite{Tang:2024dkz}. In the T-matrix approach one calculates the in-medium scattering amplitude of heavy quark and anti-quark starting from medium-modified off-shell heavy quarks (anti-quarks) and non-perturbative interaction kernel. The non-perturbative parameters of the T-matrix approach are fixed through the vacuum quarkonium spectroscopy and comparison with lattice results on charm susceptibilities, free energy of static quarks, and Wilson line correlation functions \cite{Liu:2017qah,Liu:2021rjf,Tang:2023lcn,Tang:2023tkm}. Furthermore, the spectral functions corresponding
to wave-function-optimized operators in the T-matrix approach can be described by a Lorentzian
form around the peak, and decay exponentially far away from the peak \cite{Tang:2024dkz}. As an example, in 
\autoref{fig:spf-example} we show the spectral function corresponding to the wave-function-optimized correlation function of the $\chi_b$(1P) state at $T=182$ MeV from the T-matrix calculations. The spectral function of a static quark anti-quark pair at high temperature calculated using
hard thermal loop perturbation theory also has a similar form~\cite{Burnier:2013fca}. Thus, the quasi-particle peak in the spectral function can be described by the following form:
\begin{equation}
\rho_{\rm med}(\omega,T)=\frac{1}{\pi} \frac{\Gamma(\omega)}{(\omega-M)^2+\Gamma^2(\omega)}.
\label{eq:gen-Lorentzian}
\end{equation}
For $\omega \simeq M$, the width function, $\Gamma(\omega)$ can be approximated by a constant, $\Gamma(\omega) \simeq \Gamma_0$, but should vanish quickly for $\omega$ very different from $M$. We can 
write 
\begin{equation}
\Gamma(\omega)=\Gamma_0 \Theta_{\delta}({cut}-(\omega-M))\Theta_{\delta}( {cut}+(\omega-M)),
\label{eq:gen-width}
\end{equation}
where $\Theta_{\delta}({ cut}\pm(\omega-M))$ ensures exponential decay of 
$\Gamma$ and thus of the spectral function for $\omega < M- {cut}$ and $\omega>M+ {cut}$. 
The parameter $\Gamma_0$ can be interpreted as in-medium bottomonium width.
If the exponential decay is very fast, we can neglect the exponentially small parts of
the spectral function and approximate it as
\begin{equation}
\rho_{\rm med}(\omega,T)=\frac{1}{\pi} \frac{\Gamma_0}{(\omega-M)^2+\Gamma_0^2} \Theta({cut}-|\omega-M|),
\label{eq:cut-Lorentzian}
\end{equation}
with $\Theta$ being the Heaviside step function. We call this form of the spectral
function the cut Lorentzian form. Such a form of the spectral function has been used in
the analysis of the static quark-antiquark correlators at nonzero temperatures~\cite{Bazavov:2023dci}.
It was shown that this form of the spectral function also leads to a nearly linear behavior of the continuum-subtracted effective masses in $\tau$~\cite{Bazavov:2023dci}.
In practice, a possible parametrization of $\Theta_{\delta}$ could be the following:
\begin{equation}
\Theta_{\delta}(x)=\frac{1}{2} \left[1+\mathrm{sgn}(x)\tanh(\frac{x}{\delta})\right],
\label{eq:smooth-theta}
\end{equation}
which for  very large and very small $x$ behaves as $\exp(-2x/\delta)$.  In the limit $\delta \rightarrow 0$, we recover the cut Lorentzian form.
The parametrization of the spectral function given by \autoref{eq:gen-Lorentzian}, \autoref{eq:gen-width}, and \autoref{eq:smooth-theta}, which we will call it smooth-cut Lorentizan, can describe the spectral function obtained in the T-matrix approach~\cite{Tang:2024dkz}. This is shown in \autoref{fig:spf-example} for a 1P bottomonium state as an example. The parameter $\Gamma_0$ was obtained as the width at half maximum of the T-matrix spectral function, and the parameters $cut$ and $\delta$ were adjusted to obtain the best possible description of the T-matrix spectral functions.
We also find that the Lorentzian form can describe the T-matrix spectral function very well in the interval $[M-{cut}, M+{cut}]$, where ${cut} \simeq 4\Gamma_0$. 
A similar statement holds true for the spectral function of the static quark-antiquark pair obtained in hard thermal loop perturbation theory~\cite{Burnier:2013fca}.
\begin{figure}[!htbp]
\centering
	\subfloat[The in-medium widths for the 1S (left), 2S (middle), and 3S (right) states of $\Upsilon$.]{
	\includegraphics[width=0.3\textwidth]{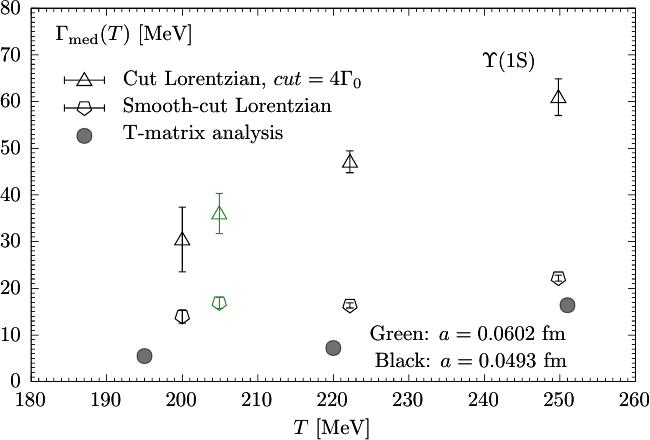}
    \quad
    \includegraphics[width=0.3\textwidth]{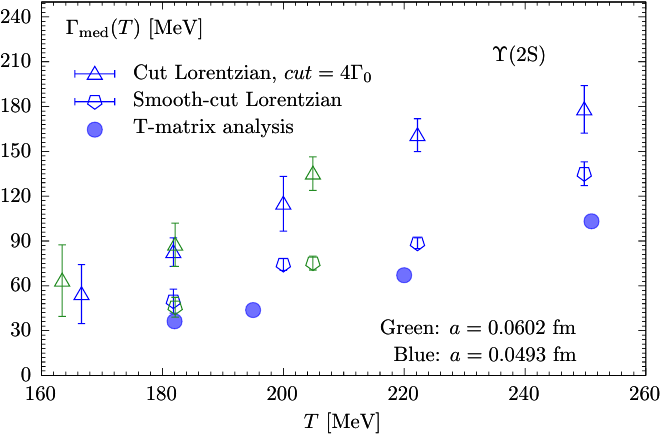}
    \quad
    \includegraphics[width=0.3\textwidth]{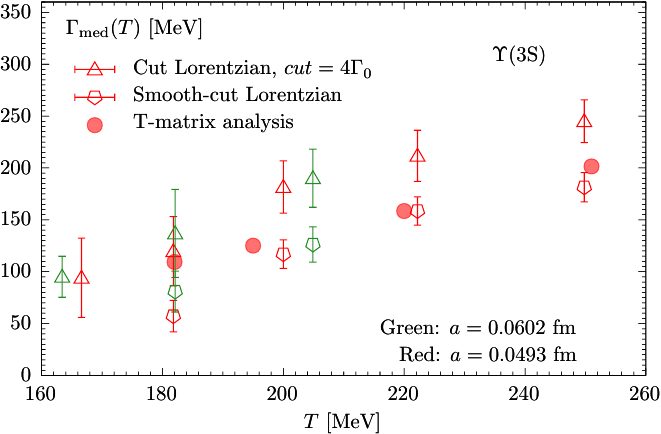}
	}\\
	\subfloat[The in-medium widths for the 1P (left), 2P (middle), and 3P (right) states of $\chi_{b0}$.]{
	\includegraphics[width=0.3\textwidth]{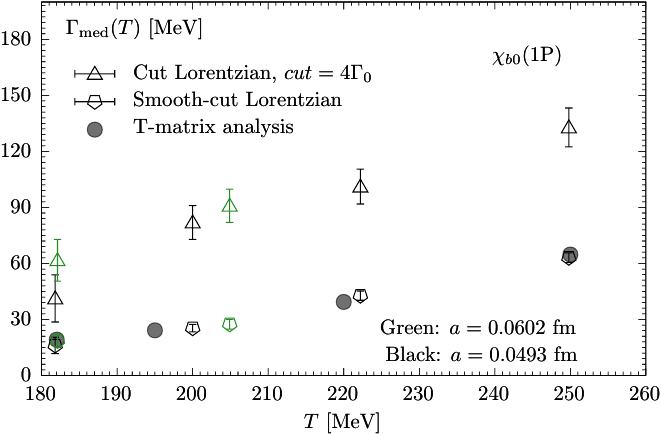}
    \quad
    \includegraphics[width=0.3\textwidth]{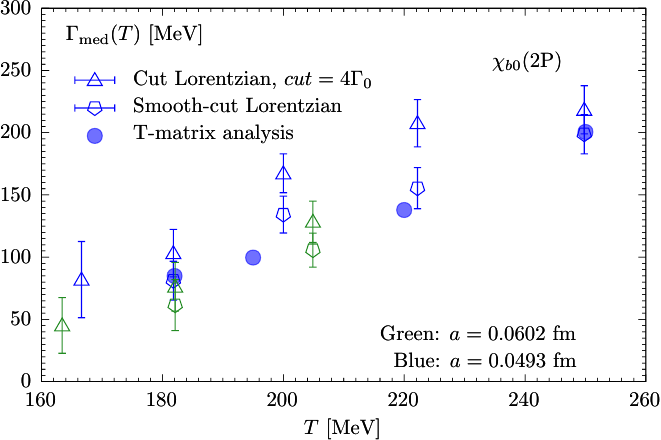}
    \quad
    \includegraphics[width=0.3\textwidth]{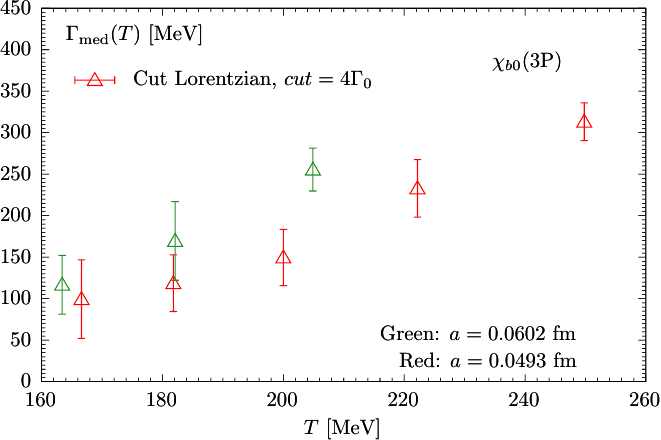}
	} 
	\caption{Temperature dependence of in-medium widths for the $\Upsilon$ and $\chi_{b0}$, given by cut Lorentzian fits with $cut=4\Gamma_0$ and smooth-cut Lorentzian fits. Filled points from the T-matrix analysis are shown for comparison. Green points are results with spacing equal to 0.0602 fm, while others correspond to 0.0493 fm.}
    \label{fig:inmedium-width}
\end{figure}

Therefore, we assume that the dominant peak can be parametrized by cut Lorentzian or smooth-cut Lorentzian forms. The contribution well below the dominant peak will be parametrized by a single delta function, since a more complicated form will lead to over-fitting of our data. Thus, we write
\begin{align}
	\rho_{\mathrm{med}}(\omega, T) 
	=A_{\mathrm{low}}(T) \delta\left(\omega-\omega_{\mathrm{low}}(T)\right) 
	+ A_{\mathrm{med}}(T) \frac{\Gamma(\omega,T)}{\left(\omega-M_{\mathrm{med}}(T)\right)^2+\Gamma^{2}(\omega,T)}.
	\label{eq:ansatz-inmedium-spectral-function}	
\end{align}
Here the function $\Gamma(\omega,T)$ is either given by \autoref{eq:gen-width} and \autoref{eq:smooth-theta} for smooth-cut Lorentzian with 
parameters ${cut}$ and $\delta$ fixed for each temperature and state as described above, or $\Gamma(\omega,T)= \Gamma_0 \Theta({cut}-|\omega-M_{\rm med}|)$ with ${cut}=4 \Gamma_0$ for the cut Lorentzian form. For both ans\"atze we have $A_{\mathrm{low}}(T)$, $\omega_{\mathrm{low}}(T)$, $A_{\mathrm{med}}(T)$, $M_{\mathrm{med}}$, and $\Gamma_{\mathrm{med}}(T)=\Gamma_0(T)$ as fit parameters.  The fits using the smooth-cut Lorentzian and the cut Lorentzian form
work  well for the continuum-subtracted effective masses, as shown in \autoref{fig:subtracted-meff-1S1P} and \autoref{fig:subtracted-meff-2S3S2P3P}. Since, as mentioned above, the continuum-subtracted effective masses are well described by constants at low temperatures, we only use the ans\"atze given by \autoref{eq:ansatz-inmedium-spectral-function} for $T>160$ MeV, specifically for $T>190$ MeV in the case of $\Upsilon$(1S) state and $T>180$ MeV for $\chi_{b0}$(1P) state.

The in-medium mass shift $\Delta M(T)$ from wave-function-optimized operators is shown in \autoref{fig:inmedium-mass-shift}.
Interestingly, we see no mass shift within estimated errors
for both the smooth-cut Lorentzian form and the cut Lorentzian form, for both lattice spacings.
This is consistent with the previous findings in Refs.~\cite{Larsen:2019bwy,Larsen:2019zqv}
and also means that the absence of the mass shift is a robust result independent from
the detailed form of the spectral function. We elaborate on this in Appendix \ref{appendix:inmedium-parameters-extraction}.
The $\Delta M(T)$ for $\Upsilon$(1S) and $\chi_{b0}$(1P) states obtained from the correlators using Gaussian sources agree with the ones shown in \autoref{fig:inmedium-mass-shift} within errors.

In \autoref{fig:inmedium-width} we show the in-medium (thermal) width, $\Gamma_{\rm med}(T)$, obtained from the correlators with wave-function-optimized operators as a function of the temperature using the cut Lorentzian and smooth-cut Lorentzian forms of the spectral function for
different bottomonium states. The thermal width obtained using the two different lattice spacings 
agree within the estimated errors, except for the $\chi_{b0}$(2P) and $\chi_{b0}$(3P) states, where we see about two-sigma deviations around the temperature of 200 MeV.

The in-medium width increases with the temperature and is larger
for higher-lying states, i.e., it follows a hierarchy pattern corresponding to the hierarchy of bottomonium masses and sizes. In the case of the $\Upsilon$(1S), $\Upsilon$(2S), and $\chi_{b0}$(1P) states, the estimates of the thermal width obtained using the cut Lorentzian 
are significantly larger than the ones obtained using the smooth-cut Lorentzian. The difference
is the largest for the $\Upsilon$(1S) state, where it is as large as a factor three. The difference
in $\Gamma_{\rm med}(T)$ obtained from the two forms of the spectral function implies that the tail
structure of the quasi-particle peak plays an important role in the observed decrease in the continuum-subtracted effective mass.
The smaller the width, the larger the contribution from the tails of the dominant peak to
the decrease in the effective masses. For this reason, the difference in $\Gamma_{\rm med}$
from using the two ans\"atze is the largest for the $\Upsilon$(1S). For the $\Upsilon$(3S) and the
$\chi_{b0}$(2P), the difference in thermal width arising from the use of the two ans\"atze 
is much smaller, and for the $\chi_{b0}$(2P) state it is almost statistically insignificant.
The reason for this is the following. When the width is large, the decrease in the effective
masses is largely determined by the behavior of the spectral function around the peak,
as explained in the Appendix \ref{appendix:inmedium-parameters-extraction}, and the tails
of the quasi-particle peak do not contribute much to the observed decrease of the continuum-subtracted effective masses. The thermal widths obtained in this work are significantly smaller than the
ones obtained in Refs. \cite{Larsen:2019bwy,Larsen:2019zqv}. The reason for this is the use
of a simple Gaussian form for the quasi-particle peak in those studies, which is not physically
motivated. For a Gaussian form of the peak, the slope of the continuum-subtracted effective 
masses is entirely determined by the width parameter. As discussed above, this is
not the case for realistic spectral functions, for which the shape of the spectral
function far away from the peak position can largely contribute to the slope of the continuum-subtracted effective masses. 

The thermal widths of $\Upsilon$(1S) and $\chi_{b0}$(1P) states obtained from the correlators 
with Gaussian sources agree with the ones shown in 
\autoref{fig:inmedium-width}. Thus, the differences in the effective masses between the Gaussian
sources and wave-function-optimized sources seen in \autoref{fig:meff-T-1S1P} can be described
in terms of $A_{\rm low}$ and $\omega_{\mathrm{low}}$, and do not lead to different values of
in-medium masses and widths.

In \autoref{fig:inmedium-width} we also compare our results for the in-medium width with the calculations using the T-matrix approach. As one can see from the figure, for the $\chi_{b0}$(1P), $\chi_{b0}$(2P),
and $\Upsilon$(3S) states, our results for $\Gamma_{\rm med}(T)$ obtained from the smooth-cut
Lorentzian agree well with the T-matrix results. At the same time, our results for the in-medium
width obtained from the smooth-cut Lorentzian fits are larger for the $\Upsilon$(1S) and $\Upsilon$(2S) states.

\section{Conclusions}
\label{sec:concl}
In this paper we studied the in-medium properties of bottomonium using NRQCD 
with extended meson operators at two lattice spacings, $a=0.0602$ fm and
$a=0.0493$ fm. We used temporal lattice extents $N_{\tau}=30-16$, covering
the temperature range $T=133-251$ MeV. By studying the temperature dependence
of bottomonium correlation functions, as well as through the comparison
of the finite-temperature correlation function to the zero-temperature correlation function,
we proposed a physically motivated parametrization of the in-medium spectral function
that contains a dominant quasi-particle peak corresponding to the bottomonium states
of interest. We found that all bottomonium states below the open-beauty threshold
can exist as well-defined quasi-particle excitations up to $T=251$ MeV, though
some of these excitations have large thermal width. For temperature $T\lesssim 167$ MeV
the existence of bottomonium states and the values of their masses can be established
without relying on any specific form of the spectral function. For the $\Upsilon$(1S) this
is true even for $T\lesssim 182$ MeV.
To obtain the in-medium bottomonium 
mass and width at higher temperatures,  we considered several parameterizations of the dominant quasi-particle peak.
We found that, irrespective of the details of these parameterizations, the in-medium 
bottomonium masses agree with the corresponding vacuum masses within errors. 
This is in contrast to the usual picture of bottomonium in quark-gluon plasma based
on color screening, where there is significant shift in bottomonium masses and
higher excited states are melted at temperatures somewhat above the crossover temperature.
The remnants of confining interactions above the crossover temperature would be consistent
with the existence of open charm hadrons suggested by the lattice QCD calculations of the generalized charm susceptibilities~\cite{Mukherjee:2015mxc,Bazavov:2023xzm}. Furthermore, a recent lattice QCD study showed that topological objects underlying confinement exist well above the crossover temperature \cite{Mickley:2024vkm}. In the future it would be interesting to extend these studies to
higher temperatures and see the eventual melting of bottomonium states. This, however, should either involve the use of anisotropic lattices
or relativistic formulations of the bottom quark since in present setup the condition $a M_b >0.9$ needed
for NRQCD cannot be ensured for high temperatures.

The estimates of the thermal width are quite sensitive to the detailed 
spectral shape of the quasi-particle peak. We considered two physically motivated
choices of this spectral shape to obtain the thermal widths, which are shown in 
\autoref{fig:inmedium-width}. These estimates of the thermal width are considerably smaller
than the previous estimates from lattice NRQCD calculations that used a Gaussian form
of the dominant peak \cite{Larsen:2019zqv,Larsen:2019bwy}, especially for the $\Upsilon$(1S).
However, the two estimates of the thermal widths differ significantly, in particular
for the $\Upsilon$(1S) state, where the differences can reach up to a factor of three. For the excited states, the differences in the thermal width obtained from the different spectral shapes
are smaller, but, in general, estimating the thermal bottomonium width on the lattice
better than a factor of two appears to be very challenging. Nevertheless, it would be very interesting
to see how our estimates of the thermal width impact the phenomenological description of the nuclear suppression
factor, $R_{AA}$, of the $\Upsilon$(1S), $\Upsilon$(2S), and $\Upsilon$(3S) states in heavy-ion collisions.

\acknowledgments
This work is supported partly by the National Key Research and Development Program of China under Contract No. 2022YFA1604900; the National Natural Science Foundation of China under Grants No.~12293064, No.~12293060 and No.~12325508, and by the U.S. Department of Energy, Office of Science, Office of Nuclear Physics through Contract No.~DE-SC0012704, through Heavy Flavor Theory for QCD Matter (HEFTY) topical collaboration in Nuclear Theory,  and within the framework of Scientific Discovery through Advance Computing (SciDAC) award Fundamental Nuclear Physics at the Exascale and Beyond. S. Meinel acknowledges support by the U.S. Department of Energy, Office of Science, Office of High Energy Physics under Award Number DE-SC0009913. W.-P. Huang acknowledges support from the Young Elite Scientists Sponsorship Program (Special Support for Doctoral Students) by CAST. R.L. was supported by the Ministry of Culture and Science of the State of Northrhine Westphalia (MKW NRW) under the funding code NW21-024-A (NRW-FAIR).

This research used awards of computer time provided by the National Energy Research Scientific Computing Center (NERSC), a U.S. Department of Energy Office of Science User Facility located at Lawrence Berkeley National Laboratory, operated under Contract No. DE-AC02- 05CH11231, 
on Frontier supercomputer in Oak Ridge Leadership Class Facility through ALCC Award, and the awards on JUWELS at GCS@FZJ, Germany and Marconi100 at CINECA, Italy. Computations for this work were carried out in part on facilities of the USQCD Collaboration, which are funded by the Office of Science of the U.S. Department of Energy.  Some of the calculations were performed on the GPU cluster in the Nuclear Science Computing Center at CCNU (NSC$^3$) and Wuhan supercomputing center. 

\appendix
\section{Details of mass extractions in vacuum\label{appendix:vacuum-mass-extraction}}
\begin{figure}[!htbp]
  \centering
    \includegraphics[width=0.45\textwidth]{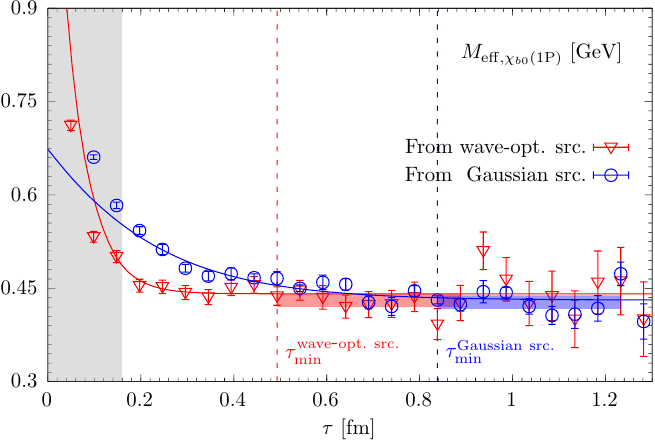}
  	\caption{The effective masses of $\chi_{b0}$(1P) from wave-function optimized (red) and Gaussian-smeared (blue) sources, enlarged from the results in \autoref{fig:meff-zero-temperature}. Bands (color-matched to the effective masses) indicate the masses extracted from single-exponential fits to the corresponding correlators. To determine the lower bound $\tau_{\text{min}}$ of the fit, initially, a two-state fit (colored curves) to the correlators is performed outside the gray-shaded area, terminating at $\tau_{\text{max}}$.  The determined intervals $[\tau_{\text{min}},\tau_{\text{max}}]$ for the single-exponential fits are represented by the span of the colored bands.}
	\label{fig:ground-fit-range}
\end{figure}

In this Appendix, we describe the detailed procedure of fitting the vacuum correlators with an exponential ans\"atze to extract the corresponding mass parameters. We fitted the correlators to a single-exponential form within a range where the ground-state contribution dominates. Thus, careful selection of the fit interval is essential to eliminate contamination from higher states while maintaining statistical precision. To achieve this, we followed a procedure adapted from Refs.~\cite{Fritzsch:2012wq,Bahr2015Form}. This procedure is exemplified in \autoref{fig:ground-fit-range}, using the $\chi_{b0}$(1P) case as an illustration. The upper end of the fit interval $\tau_{\text{max}}$ is determined based on the temporal separation where the signal-to-noise ratio of the effective mass $\meff(\tau)$ reaches a specified threshold, with the effective mass defined as $\meff(\tau)=-\partial_{\tau} \log C(\tau)$. In our analyses, we set a conservative criterion of $1\%$ for measurements employing Gaussian-smearing sources and $8\%$ for those using wave-function-optimized sources.

To determine the lower end of the fit interval $\tau_{\text{min}}$, we initially performed a two-exponential fit to the correlators within a range where the two-state ans\"atze $f_n(\tau)=\sum_{i=0}^{n} A_i e^{-E_i \tau}$ with $n=1$ described the data well. This is illustrated by the solid colored lines in \autoref{fig:ground-fit-range}, with the two-state fits performed outside the gray-shaded area and ending at $\tau_{\text{max}}$. The fit parameters $A_{n=0,1}$ and $E_{n=0,1}$ were then used to reconstruct the excited-state contribution to the effective mass as $\delta M(\tau) = \log[f_1(\tau)/f_1(\tau+a)] - \log[f_0(\tau)/f_0(\tau+a)]$. We set $\tau_{\text{min}}$ to the smallest $\tau$ value where $\delta M(\tau) / E_0 < 25\% \times \delta_{\meff}(\tau) / \meff(\tau)$ is satisfied, with $\delta_{\meff}(\tau)$ representing the statistical uncertainty of $\meff$ at $\tau$. In this way, from $\tau_{\text{min}}$, the excited-state contamination (or, equivalently, the systematic uncertainty due to the fit ans\"atze) falls below the statistical uncertainty.

For the $\chi_{b0}$(1P) case, $\tau_{\text{min}}$ is indicated by the dashed lines for wave-function optimized and Gaussian-smeared measurements respectively in \autoref{fig:ground-fit-range}, with the intervals $[\tau_{\text{min}},\tau_{\text{max}}]$ marked as the span of the colored bands. Within these intervals, the ground state dominates, allowing for a single-exponential fit to extract the information of the ground state. The error of the extracted mass is estimated using the Bootstrap method. The final central value of the mass was the median of the Bootstrap ensemble of fit results, with the error quoted as the 1-$\sigma$ width of this distribution. It is worth mentioning that for quantities such as mass differences, which depend on results from different fits, each fit was performed on the same Bootstrap sample. The final quantity for one sample is then computed using these fit results from the same sample, giving an approximate probability distribution. The mean value and the 1-$\sigma$ width of this distribution are then quoted.

\section{Extractions of in-medium parameters at finite temperatures\label{appendix:inmedium-parameters-extraction}}
In this Appendix, details of the fitting procedure to extract the in-medium bottomonium parameters are presented, as well as the consequences of different parametrizations of the in-medium spectral function $\rho_{\rm med}(\omega, T)$. For illustration, we will mostly focus on the $\Upsilon$ case as the situations in the other cases are very similar.

As discusssed in \autoref{sec:in-med} of the main text, the in-medium bottomonium parameters such as the widths and masses are encoded in the continuum-subtracted in-medium spectral function $\rho_{\rm med}(\omega, T)$. Furthermore, $\rho_{\rm med}(\omega, T)$ includes a quasi-particle peak containing information on the in-medium bottomonium masses and widths, and also a small contribution at frequencies well below the dominant peak controlling the behavior of the correlator at large $\tau$ around $1/T$. For the first part, we use the physically motivated cut Lorentzian ans\"atze or the smooth-cut Lorentzian ans\"atze, as discussed in the main text, while for the other part at $\omega$ well below the dominant-peak position, a single delta function  is sufficient to describe the correlators at large $\tau$, shown as the first term of \autoref{eq:ansatz-inmedium-spectral-function}. In the main text it was argued that, for the cut Lorentzian, a reasonable choice for the
cut is $4 \Gamma_{0}$. Then an important question to address is how the obtained width depends on the choice of the parameter $cut$. A smaller value of $cut$ will result in a larger width. On the other hand the value of $cut$ cannot be too small because that would mean that the quasi-particle peak is not well defined, as other parts the spectral function provide a much larger contribution to the correlator. The value ${cut}=2 \Gamma_{0}$ is probably the smallest value 
compatible with the existence of a quasi-particle peak, as the spectral function around the quasi-particle peak will decrease by a large factor before other structure in
the spectral function becomes dominant. Therefore, we also considered the choice ${cut}=2\Gamma_{0}$ in our analysis as discussed below. 
For sufficiently low temperature, the thermal width can be too small to be resolved, given our statistical errors. In such cases, it makes sense to
set the thermal width to zero, i.e., to assume a delta function for the quasi-particle peak. 
For sufficiently low temperature, the small contribution below the quasi-particle peak also does
not play a role and can be neglected. 
This corresponds to fitting the effective masses by
a constant. The results of these fits are shown in \autoref{tab:fit-constant-parameters-Upsilon} and \autoref{tab:fit-constant-parameters-chi0} for the
$\Upsilon$ and $\chi_{b0}$ states respectively. 
In the cases of the $\Upsilon$(1S) state, the constant fit has worse $\chi^2/d.o.f$ compared to
other fits only for $T \gtrsim 182$ MeV, while
for other bottomonium  states the constant fit has worse  $\chi^2/d.o.f$ only for $T\gtrsim 163$ MeV.

In \autoref{tab:fit-Lorentzian-spacing0493-parameters-Uspilon} and \autoref{tab:fit-Lorentzian-spacing0602-parameters-Uspilon} we present our fit results for the cut Lorentzian form with $cut=2\Gamma_{0}$ and $4\Gamma_{0}$ for $a=0.0493$ fm and $a=0.0602$ fm, respectively. We performed both correlated and uncorrelated fits.  The fit parameters for the correlated and uncorrelated fits agree within
errors.  We performed fits excluding the delta function contribution at small $\omega$. For these
fits, we exclude $2-4$ data points close to $\tau=1/T$ from the fits. As one can see from the tables,
excluding the delta  function does not change the value of $M_{\rm med}$.

We also performed fits using the Gaussian form of the quasi-particle peak in the spectral
function given by
\autoref{eq:Gaussian} to make contact with previous analyses \cite{Larsen:2019zqv,Larsen:2019bwy}, which
are summarized in \autoref{tab:fit-Gaussian-parameters-Uspilon}.

\begin{figure}[htbp]
\centering
  	\includegraphics[width=0.3\textwidth]{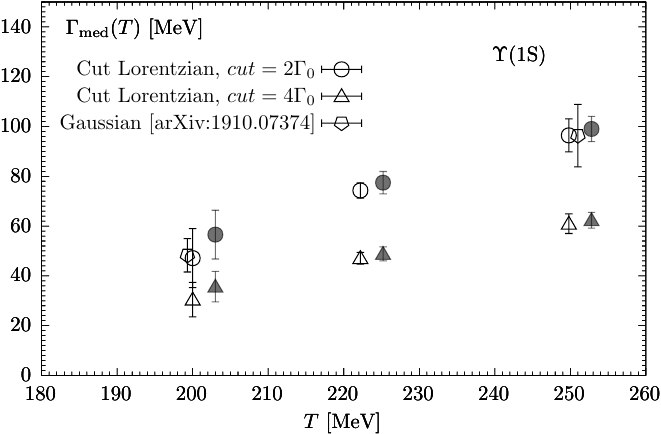}
    \quad
    \includegraphics[width=0.3\textwidth]{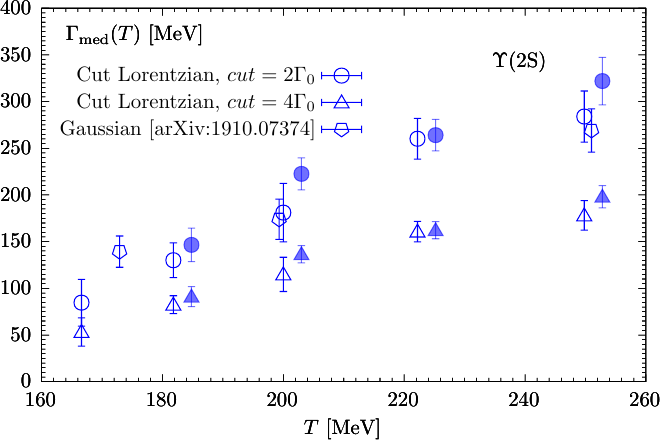}
    \quad
    \includegraphics[width=0.3\textwidth]{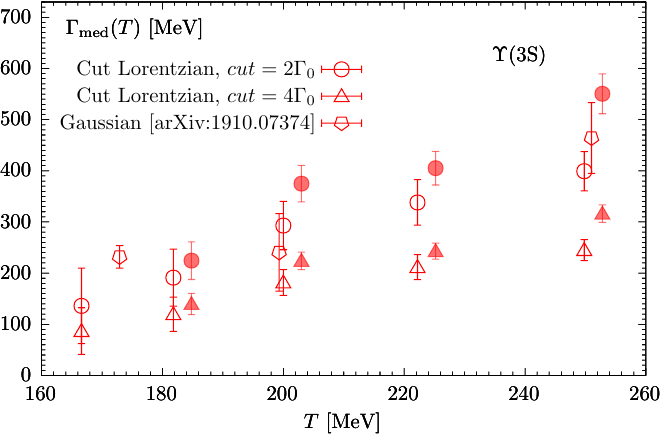}
    \includegraphics[width=0.3\textwidth]{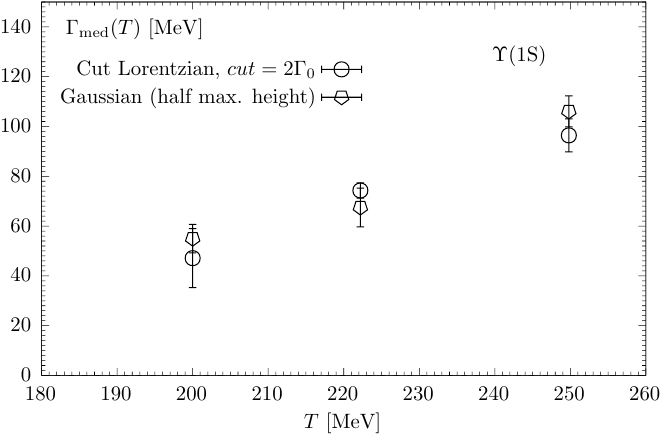}
    \quad
    \includegraphics[width=0.3\textwidth]{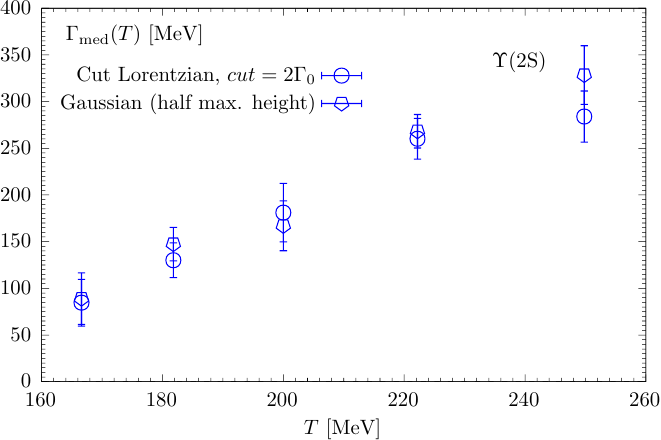}
    \quad
    \includegraphics[width=0.3\textwidth]{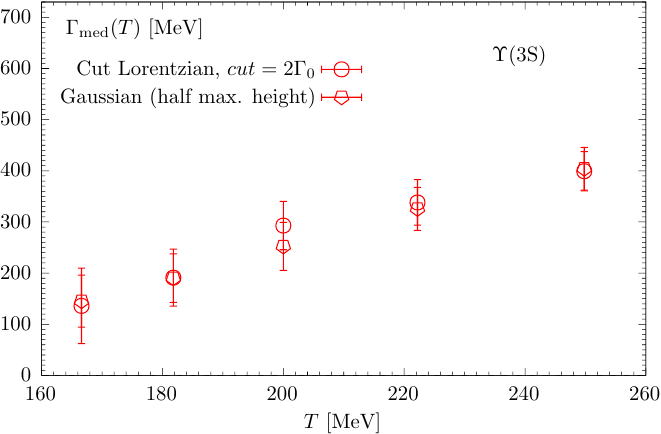}
    \caption{Comparisons of in-medium widths between different types of ans\"atze. Open points for $T>180$ MeV are from a given ans\"atze including a delta function term, while open ones at $T<180$ MeV are given by the ans\"atze excluding a delta function term.
    Filled points (slightly shifted for visibility) represent results from the corresponding ansatz excluding the delta function term at $T>180$ MeV, used to assess the effect of excluding the delta function. Pentagon points in the top panel are from previous similar work but extracting width using Gaussian ans\"atze, while these in the bottom panel are from Gaussian fits on the continuum-subtracted correlators of this work.}
    \label{fig:appx-different-fits-width}
\end{figure}

\begin{figure}[htbp]
\centering
  	\includegraphics[width=0.45\textwidth]{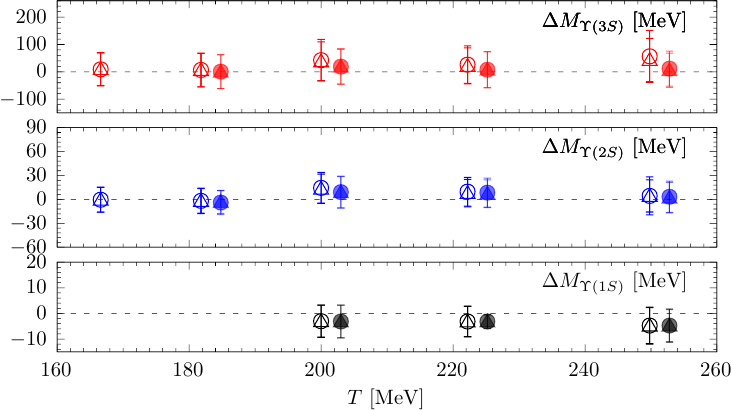}
    \caption{Comparisons of in-medium mass shifts between different types of ans\"atze (same symbols as the top panel of \autoref{fig:appx-different-fits-width}). Open points for $T>180$ MeV are from a given ans\"atze including a delta function term, while open ones at $T<180$ MeV are given by the ans\"atze excluding a delta function term. Filled points (slightly shifted for visibility) represent results from the corresponding ansatz excluding the delta function term at $T>180$ MeV, used to assess the effect of excluding the delta function.}
    \label{fig:appx-different-fits-mass}
\end{figure}

The in-medium widths and mass shifts from different types of parametrizations of the in-medium spectral function $\rho_{\rm med}(\omega, T)$ are shown in \autoref{fig:appx-different-fits-width} and \autoref{fig:appx-different-fits-mass}. It can be seen from the comparison of open and filled points with the same shape in \autoref{fig:appx-different-fits-width} that, for the excited bottomonium states, the widths extracted using the ans\"atze excluding a delta-function term tend to be systematically larger than those including the delta function, no matter what type of ans\"atze is used. This is especially obvious in the cases of high temperatures and for excited states. This indicates that the minor structure below the dominant peak of the spectral function matters for these cases, as expected. This delta peak is not well separated from the main peak structure of the spectral function, and interferences between them are possible. Thus, the tail of the main peak will compensate for the delta peak contribution if the delta term is excluded during the fits, leading to a larger width. In the case of the Gaussian fits the thermal width in this is defined as the width at the half maximum, which is related to
the Gaussian width parameters as $\sqrt{2 \ln 2} \Gamma_G$ \cite{Shi:2021qri}.
With this definition the thermal width from Gaussian fits width  agrees with the cut Lorentzian results with $cut=2\Gamma_0$. As the cut increases to $4\Gamma_0$, the width extracted from cut Lorentzian ans\"atze decreases to about 50\% $\sim$ 60\%  of the $cut=2\Gamma_0$ result, also stressing the importance of modeling the tail of the main peak structure of the spectral function. 
We also compare our estimates for the thermal width obtained from the cut Lorentzian form to estimates using the Gaussian ans\"atze for the dominant peak obtained in this analysis as well
as in Ref. \cite{Larsen:2019bwy}.
Here again the width at half maximum was used to define
the thermal width. Since the choice $cut=2 \Gamma_{0}$ is the smallest value of cut still compatible with a well-defined
quasi-particle peak and leads to the larger thermal width, the estimate of the width obtained from the
Gaussian form of the dominant peak can be considered as an upper bound on the thermal bottomonium width.
The in-medium mass shifts in \autoref{fig:appx-different-fits-mass} are all consistent with zero for all states under our consideration, irrelevant of the parametrizations of $\rho_{\rm med}(\omega, T)$.

The detailed choice of the quasi-particle peak has significant effect on the estimated thermal 
bottomonium width, while the position of the peak turns out to be robust. 
Then the question arises which other features of the spectral functions are robust and not sensitive to
the details of the spectral shape of the quasi-particle peak.
In Ref. \cite{Bazavov:2023dci} it was argued that the cumulants  of the continuum-subtracted spectral function, defined
as
\begin{equation}
c_1=\langle \omega \rangle, 
c_2=\langle \omega^2 \rangle -\langle \omega \rangle^2,
\text{~with~}
\langle \dots \rangle = \int_{-\infty}^{\infty} \mathrm{d} \omega ~ \rho_{\rm med}(\omega,T)\dots\,,
\end{equation}
provide a robust characterization of the spectral function given the lattice data.
The cumulants of the spectral function can be related to the cumulants of the correlation function
for small $\tau$ \cite{Bazavov:2023dci}. In particular, $c_2$ is related to the second cumulant of
the correlation function at $\tau=0$ \cite{Bazavov:2023dci}, which is the slope of the effective mass
at small $\tau$. In the case of the Gaussian form of the dominant peak, $c_2=\Gamma_G$.
In \autoref{tab:fit-Lorentzian-spacing0493-parameters-Uspilon}, \autoref{tab:fit-Lorentzian-spacing0602-parameters-Uspilon} and \autoref{tab:fit-Gaussian-parameters-Uspilon}, we show the values of $\sqrt{c_2}$
in lattice units. We see from the tables that, once the delta function at small $\omega$ is included
in the fit, the values of $c_2$ obtained from the fit using the cut Lorentzian with $cut=2 \Gamma_{0}$ and $cut=4 \Gamma_{0}$, and using the Gaussian form agree. 
Thus, our results for $c_2$ can be used
to constrain the parameters of other possible models for the spectral function and provide an alternative estimate of the thermal width.

\begin{table}[htbp]
\centering
\resizebox{0.8\textwidth}{!}{
\begin{tabular}{*{5}{c}}
\hline
\hline
\multirow{2}{*}{}                
&\multicolumn{4}{c}{[$aM_{\text{med}}$, ${\chi^2}/{\text{d.o.f.}}$] for $a$=0.0602 fm}   
\\
\cline{2-5} 
&
$T=$136.6 MeV&
163.4 MeV&
182.1 MeV&
204.9 MeV
\\
$\Upsilon$(1S) 
& [0.515(1), 0.85] & [0.515(1), 0.70] & [0.515(1), 1.01] & [0.512(1), 1.82]
\\
$\Upsilon$(2S) 
& [0.689(2), 1.37] & [0.685(1), 1.26] & [0.683(1), 2.89] & [0.672(1), 8.26]
\\
$\Upsilon$(3S) 
& [0.796(7), 1.03] & [0.788(6), 1.02] & [0.780(6), 2.46] & [0.756(6), 9.68]
\\
\hline\hline
\multirow{2}{*}{}                
&\multicolumn{4}{c}{[$aM_{\text{med}}$, ${\chi^2}/{\text{d.o.f.}}$] for $a$=0.0493 fm } 
\\
\cline{2-5} 
&
$T=$133.3 MeV&
142.8 MeV&
153.8 MeV&
166.6 MeV
\\
$\Upsilon$(1S) 
& [0.615(1), 1.56]
& [0.614(1), 1.27]
& [0.614(1), 1.06]
& [0.614(1), 1.03]
\\
$\Upsilon$(2S) 
& [0.755(1), 1.28]
& [0.756(1), 1.05]
& [0.755(1), 1.64]
& [0.754(1), 1.24]
\\
$\Upsilon$(3S) 
& [0.842(5), 0.97]
& [0.846(5), 1.09]
& [0.843(4), 1.08]
& [0.839(5), 1.25]
\\
\hline
\multirow{2}{*}{}                
&\multicolumn{4}{c}{[$aM_{\text{med}}$, ${\chi^2}/{\text{d.o.f.}}$] for $a$=0.0493 fm } 
\\
\cline{2-5} 
&      
$T=$181.8 MeV&
200.0 MeV&
222.2 MeV&
249.8 MeV
\\
$\Upsilon$(1S) 
& [0.613(1), 0.97]
& [0.613(1), 1.30]
& [0.611(1), 2.13]
& [0.610(1), 6.25]
\\
$\Upsilon$(2S)
& [0.749(1), 3.31]
& [0.745(1), 4.06]
& [0.731(1), 18.79]
& [0.720(1), 36.83]
\\
$\Upsilon$(3S) 
& [0.831(5), 1.99]
& [0.808(5), 6.05]
& [0.794(4), 23.69]
& [0.768(4), 58.34]
\\
\hline\hline
\end{tabular}
}
\caption{Summary of parameters from constant fits on continuum-subtracted effective masses, for $\Upsilon$ cases.}
\label{tab:fit-constant-parameters-Upsilon}
\end{table}

\begin{table}[htbp]
\centering
\resizebox{0.8\textwidth}{!}{
\begin{tabular}{*{5}{c}}
\hline\hline
\multirow{2}{*}{}                
&\multicolumn{4}{c}{[$aM_{\text{med}}$, ${\chi^2}/{\text{d.o.f.}}$] for $a$=0.0602 fm}   
\\
\cline{2-5} 
&
$T=$136.6 MeV&
163.4 MeV&
182.1 MeV&
204.9 MeV
\\
$\chi_{b0}$(1P)
& [0.647(2), 1.23]
& [0.644(2), 1.57]
& [0.642(2), 2.41]
& [0.626(1), 1.84]
\\
$\chi_{b0}$(2P)
& [0.751(2), 1.18]
& [0.747(2), 1.17]
& [0.744(2), 2.47]
& [0.732(2), 6.42]
\\
$\chi_{b0}$(3P)
& [0.847(13), 0.65]
& [0.842(12), 0.61]
& [0.830(13), 1.81]
& [0.798(11), 6.26]
\\
\hline\hline
\multirow{2}{*}{}                
&\multicolumn{4}{c}{[$aM_{\text{med}}$, ${\chi^2}/{\text{d.o.f.}}$] for $a$=0.0493 fm } 
\\
\cline{2-5} 
&
$T=$133.3 MeV&
142.8 MeV&
153.8 MeV&
166.6 MeV
\\
$\chi_{b0}$(1P)
& [0.718(1), 1.20]
& [0.718(1), 0.69]
& [0.717(1), 1.20]
& [0.717(1), 0.98]
\\
$\chi_{b0}$(2P)
& [0.809(4), 1.08]
& [0.810(4), 0.98]
& [0.809(4), 1.08]
& [0.807(4), 0.91]
\\
$\chi_{b0}$(3P)
& [0.891(9), 0.88]
& [0.892(7), 0.63]
& [0.890(7), 0.94]
& [0.888(8), 1.07]
\\
\hline
\multirow{2}{*}{}                
&\multicolumn{4}{c}{[$aM_{\text{med}}$, ${\chi^2}/{\text{d.o.f.}}$] for $a$=0.0493 fm } 
\\
\cline{2-5} 
&      
$T=$181.8 MeV&
200.0 MeV&
222.2 MeV&
249.8 MeV
\\
$\chi_{b0}$(1P)
& [0.714(1), 1.89]
& [0.711(1), 1.56]
& [0.705(1), 5.90]
& [0.698(1), 16.90]
\\
$\chi_{b0}$(2P)
& [0.800(4), 1.72]
& [0.791(4), 4.01]
& [0.772(4), 18.65]
& [0.751(4), 43.25]
\\
$\chi_{b0}$(3P)
& [0.876(8), 2.06]
& [0.863(7), 5.40]
& [0.833(7), 19.26]
& [0.791(6), 53.58]
\\
\hline\hline
\end{tabular}
}
\caption{Summary of parameters from constant fits on continuum-subtracted effective masses, for $\chi_{b0}$ cases.}
\label{tab:fit-constant-parameters-chi0}
\end{table}

\begin{table}[htbp]
\centering
\resizebox{\textwidth}{!}{
\begin{tabular}{*{17}{c}}
\hline\hline
\multirow{3}{*}{$a$}                                &
\multirow{3}{*}{$T$}                                &
\multirow{3}{*}{State}                              &
\multicolumn{6}{c}{Cut Lorentzian uncorrelated fits with cut=$2\Gamma_0$} &
\multicolumn{6}{c}{Cut Lorentzian correlated fits with cut=$2\Gamma_0$} &       
\\  
& & & \multicolumn{3}{c}{Excl. $\delta$ term} &
      \multicolumn{3}{c}{Incl. $\delta$ term} &
      \multicolumn{3}{c}{Excl. $\delta$ term} &
      \multicolumn{3}{c}{Incl. $\delta$ term} 
\\
\cline{4-6} 
\cline{7-9} 
\cline{10-12}
\cline{13-15} 
[fm] & [MeV] &
&$aM_{\text{med}}$ & $a\sqrt{c_2}$ & \tiny $\frac{\chi^2}{\text{d.o.f.}}$
&$aM_{\text{med}}$ & $a\sqrt{c_2}$ & \tiny $\frac{\chi^2}{\text{d.o.f.}}$
&$aM_{\text{med}}$ & $a\sqrt{c_2}$ & \tiny $\frac{\chi^2}{\text{d.o.f.}}$
&$aM_{\text{med}}$ & $a\sqrt{c_2}$ & \tiny $\frac{\chi^2}{\text{d.o.f.}}$ 
\\
\hline \hline 
\vspace{-3ex}
\multirow{15}{*}{0.0493} & & & & & & &
\\
 & \multirow{3}{*}{249.8} & $\Upsilon$(1S) 
 & 0.613(1) & 0.017(2) & 0.60
 & 0.613(1) & 0.018(7) & 0.72
 & 0.613(1) & 0.019(3) & 1.43
 & 0.613(1) & 0.018(7) & 1.45
\\
 & & $\Upsilon$(2S) 
 & 0.764(4) & 0.069(3) & 0.48
 & 0.758(5) & 0.064(10) & 0.26
 & 0.764(4) & 0.068(3) & 1.08
 & 0.758(6) & 0.063(18) & 1.43
\\
 & & $\Upsilon$(3S) 
 & 0.865(12) & 0.105(7) & 0.97
 & 0.845(11) & 0.089(11) & 0.39
 & 0.864(11) & 0.102(6) & 2.45
 & 0.846(12) & 0.089(16) & 1.34
\\
\cline{3-15} 
 & \multirow{3}{*}{222.2} & $\Upsilon$(1S) 
 & 0.613(1) & 0.013(3) & 0.64
 & 0.613(1) & 0.012(3) & 0.57
 & 0.613(1) & 0.012(3) & 1.76
 & 0.613(1) & 0.012(5) & 1.24
\\
 & & $\Upsilon$(2S) 
 & 0.760(3) & 0.054(3) & 0.43
 & 0.758(4) & 0.057(7) & 0.36
 & 0.760(3) & 0.053(3) & 1.98
 & 0.755(4) & 0.034(11) & 1.96
\\
 & & $\Upsilon$(3S)
 & 0.854(10) & 0.083(6) & 0.70
 & 0.844(12) & 0.075(13) & 0.61
 & 0.854(10) & 0.081(7) & 1.84
 & 0.835(9) & 0.037(19) & 1.28
\\
\cline{3-15} 
 & \multirow{3}{*}{200.0} & $\Upsilon$(1S) 
 & 0.614(1) & 0.010(3) & 0.51
 & 0.614(1) & 0.011(6) & 0.72
 & 0.614(1) & 0.012(3) & 1.03
 & 0.614(1) & 0.010(5) & 1.23
\\
 & & $\Upsilon$(2S)
 & 0.761(4) & 0.045(4) & 0.51
 & 0.759(4) & 0.035(12) & 0.39
 & 0.761(3) & 0.045(3) & 1.72
 & 0.760(4) & 0.039(11) & 1.75
\\
 & & $\Upsilon$(3S) 
 & 0.857(11) & 0.072(7) & 0.65
 & 0.843(11) & 0.061(13) & 0.34
 & 0.858(11) & 0.072(7) & 1.88
 & 0.844(11) & 0.062(26) & 1.67
\\
\cline{3-15} 
 & \multirow{3}{*}{181.8} & $\Upsilon$(1S) 
 & 0.614(1) & 0.003(4) & 0.47
 & 0.614(1) & 0.003(5) & 0.59
 & 0.614(1) & 0.004(4) & 0.98
 & 0.614(1) & 0.002(6) & 1.71
\\
 & & $\Upsilon$(2S) 
 & 0.757(3) & 0.029(4) & 0.61
 & 0.756(2) & 0.028(4) & 0.76
 & 0.758(2) & 0.030(3) & 0.99
 & 0.756(2) & 0.031(4) & 1.66
\\
 & & $\Upsilon$(3S) 
 & 0.846(8) & 0.045(8) & 0.30
 & 0.842(9) & 0.040(17) & 0.25
 & 0.847(8) & 0.045(7) & 1.35
 & 0.841(8) & 0.022(25) & 1.55
\\
\cline{3-15} 
 & \multirow{3}{*}{166.6} & $\Upsilon$(1S)
 & 0.615(1) & 0.005(4) & 0.94
 & 0.614(1) & 0.002(4) & 0.98
 & 0.614(1) & 0.000(5) & 1.30
 & 0.614(1) & 0.000(1) & 1.09
\\
 & & $\Upsilon$(2S) 
 & 0.757(2) & 0.019(6) & 0.64
 & 0.757(3) & 0.013(11) & 0.76
 & 0.757(2) & 0.016(7) & 0.98
 & 0.757(2) & 0.018(9) & 1.06
\\
 & & $\Upsilon$(3S) 
 & 0.846(8) & 0.031(13) & 0.45
 & 0.843(7) & 0.009(2) & 0.36
 & 0.846(8) & 0.029(12) & 1.01
 & 0.844(8) & 0.012(22) & 1.16
\\
\hline\hline
&&&
\multicolumn{6}{c}{Cut Lorentzian uncorrelated fits with cut=$4\Gamma_0$} &
\multicolumn{6}{c}{Cut Lorentzian correlated fits with cut=$4\Gamma_0$} &        
\\  
& & & \multicolumn{3}{c}{Excl. $\delta$ term} &
      \multicolumn{3}{c}{Incl. $\delta$ term} &
      \multicolumn{3}{c}{Excl. $\delta$ term} &
      \multicolumn{3}{c}{Incl. $\delta$ term} 
\\
\cline{4-6} 
\cline{7-9} 
\cline{10-12} 
\cline{13-15} 
&&
&$aM_{\text{med}}$ & $a\sqrt{c_2}$ & \tiny $\frac{\chi^2}{\text{d.o.f.}}$
&$aM_{\text{med}}$ & $a\sqrt{c_2}$ & \tiny $\frac{\chi^2}{\text{d.o.f.}}$
&$aM_{\text{med}}$ & $a\sqrt{c_2}$ & \tiny $\frac{\chi^2}{\text{d.o.f.}}$
&$aM_{\text{med}}$ & $a\sqrt{c_2}$ & \tiny $\frac{\chi^2}{\text{d.o.f.}}$ 
\\
\hline \hline 
\vspace{-3ex}
\multirow{15}{*}{0.0493} & & & & & & &
\\
 & \multirow{3}{*}{249.8} & $\Upsilon$(1S) 
 & 0.613(1) & 0.027(4) & 0.60
 & 0.613(1) & 0.018(7) & 0.79
 & 0.613(1) & 0.029(4) & 1.42
 & 0.613(1) & 0.018(7) & 1.35
\\
 & & $\Upsilon$(2S) 
 & 0.762(4) & 0.104(4) & 0.38
 & 0.757(5) & 0.064(9) & 0.25
 & 0.762(3) & 0.104(4) & 1.65
 & 0.758(6) & 0.064(18) & 1.23
\\
 & & $\Upsilon$(3S) 
 & 0.859(10) & 0.153(7) & 0.73
 & 0.844(10) & 0.086(10) & 0.38
 & 0.859(9) &  0.150(7) & 2.41
 & 0.846(10) & 0.089(15) & 1.41
\\
\cline{3-15} 
 & \multirow{3}{*}{222.2} & $\Upsilon$(1S) 
 & 0.613(1) & 0.020(4) & 0.63
 & 0.613(1) & 0.012(3) & 0.57
 & 0.613(1) & 0.019(5) & 1.76
 & 0.613(1) & 0.011(5) & 1.60
\\
 & & $\Upsilon$(2S) 
 & 0.759(3) & 0.082(4) & 0.38
 & 0.758(4) & 0.056(5) & 0.34
 & 0.759(3) & 0.081(5) & 1.79
 & 0.755(4) & 0.035(10) & 1.59
\\
 & & $\Upsilon$(3S)
 & 0.852(9) & 0.123(8) & 0.61
 & 0.844(11) & 0.075(12) & 0.60
 & 0.852(9) & 0.121(8) & 1.30
 & 0.835(9) & 0.038(19) & 1.21
\\
\cline{3-15} 
 & \multirow{3}{*}{200.0} & $\Upsilon$(1S) 
 & 0.614(1) & 0.016(4) & 0.51
 & 0.614(1) & 0.011(6) & 0.72
 & 0.614(1) & 0.018(4) & 1.03
 & 0.614(1) & 0.010(5) & 1.34
\\
 & & $\Upsilon$(2S)
 & 0.761(3) & 0.069(5) & 0.49
 & 0.759(4) & 0.036(11) & 0.39
 & 0.761(3) & 0.069(5) & 1.64
 & 0.760(4) & 0.039(11) & 1.72
\\
 & & $\Upsilon$(3S) 
 & 0.856(9) & 0.109(9) & 0.58
 & 0.843(11) & 0.061(22) & 0.34
 & 0.857(9) & 0.107(8) & 1.52
 & 0.844(11) & 0.063(26) & 1.56
\\
\cline{3-15} 
 & \multirow{3}{*}{181.8} & $\Upsilon$(1S) 
 & 0.614(1) & 0.005(5) & 0.47
 & 0.614(1) & 0.003(5) & 0.61
 & 0.614(1) & 0.007(7) & 0.98
 & 0.614(1) & 0.001(7) & 1.77
\\
 & & $\Upsilon$(2S) 
 & 0.757(2) & 0.045(6) & 0.61
 & 0.755(2) & 0.028(4) & 0.76
 & 0.757(2) & 0.047(5) & 0.98
 & 0.756(2) & 0.030(3) & 1.99
\\
 & & $\Upsilon$(3S) 
 & 0.846(8) & 0.070(11) & 0.29
 & 0.842(9) & 0.040(16) & 0.25
 & 0.847(8) & 0.070(10) & 1.30
 & 0.841(8) & 0.023(24) & 1.54
\\
\cline{3-15} 
 & \multirow{3}{*}{166.6} & $\Upsilon$(1S)
 & 0.615(1) & 0.008(6) & 0.74
 & 0.614(1) & 0.002(4) & 0.79
 & 0.614(1) & 0.000(8) & 1.30
 & 0.614(1) & 0.000(1) & 1.09
\\
 & & $\Upsilon$(2S) 
 & 0.757(2) & 0.030(10) & 0.65
 & 0.757(3) & 0.013(10) & 0.77
 & 0.757(2) & 0.026(11) & 0.98
 & 0.757(2) & 0.018(8) & 1.06
\\
 & & $\Upsilon$(3S) 
 & 0.846(8) & 0.049(19) & 0.45
 & 0.843(7) & 0.009(20) & 0.36
 & 0.846(8) & 0.046(19) & 1.01
 & 0.844(8) & 0.012(22) & 1.15
\\
\hline\hline
\end{tabular}
}
\caption{Summary of parameters from correlated and uncorrelated fits on continuum-subtracted correlators or effective masses using cut Lorentzian ans\"atze, taking $\Upsilon$ case as an example.}
\label{tab:fit-Lorentzian-spacing0493-parameters-Uspilon}
\end{table}

\begin{table}[htbp]
\centering
\resizebox{\textwidth}{!}{
\begin{tabular}{*{17}{c}}
\hline\hline
\multirow{3}{*}{$a$}                                &
\multirow{3}{*}{$T$}                                &
\multirow{3}{*}{State}                              &
\multicolumn{6}{c}{Cut Lorentzian uncorrelated fits with cut=$2\Gamma_0$} &
\multicolumn{6}{c}{Cut Lorentzian correlated fits with cut=$2\Gamma_0$} &         
\\  
& & & \multicolumn{3}{c}{Excl. $\delta$ term} &
      \multicolumn{3}{c}{Incl. $\delta$ term} &
      \multicolumn{3}{c}{Excl. $\delta$ term} &
      \multicolumn{3}{c}{Incl. $\delta$ term} 
\\
\cline{4-6} 
\cline{7-9} 
\cline{10-12} 
\cline{13-15} 
[fm] & [MeV] &
&$aM_{\text{med}}$ & $a\sqrt{c_2}$ & \tiny $\frac{\chi^2}{\text{d.o.f.}}$
&$aM_{\text{med}}$ & $a\sqrt{c_2}$ & \tiny $\frac{\chi^2}{\text{d.o.f.}}$
&$aM_{\text{med}}$ & $a\sqrt{c_2}$ & \tiny $\frac{\chi^2}{\text{d.o.f.}}$
&$aM_{\text{med}}$ & $a\sqrt{c_2}$ & \tiny $\frac{\chi^2}{\text{d.o.f.}}$ 
\\
\hline \hline 
\vspace{-3ex}
\multirow{9}{*}{0.0602}  & & & & & &
\\
 & \multirow{3}{*}{204.9} & $\Upsilon$(1S) 
 & 0.515(1) & 0.017(3) & 0.56
 & 0.515(1) & 0.015(2) & 0.79
 & 0.516(1) & 0.018(3) & 0.98
 & 0.515(1) & 0.015(3) & 1.62
\\
 & & $\Upsilon$(2S)
 & 0.694(4) & 0.056(4) & 1.52
 & 0.692(4) & 0.060(6) & 1.54
 & 0.695(5) & 0.059(5) & 2.07
 & 0.693(4) & 0.062(5) & 2.06
\\
 & & $\Upsilon$(3S) 
 & 0.805(15) & 0.087(9) & 0.84
 & 0.800(16) & 0.088(15) & 1.00
 & 0.806(15) & 0.088(10) & 1.25
 & 0.803(15) & 0.095(11) & 1.22
\\
\cline{3-15} 
 & \multirow{3}{*}{182.1} & $\Upsilon$(1S) 
 & 0.516(1) & 0.011(5) & 0.61
 & 0.516(1) & 0.012(4) & 0.55
 & 0.517(1) & 0.016(3) & 0.94
 & 0.516(1) & 0.012(4) & 1.28
 \\
 & & $\Upsilon$(2S) 
 & 0.692(4) & 0.036(7) & 1.16
 & 0.691(4) & 0.038(6) & 1.13
 & 0.694(4) & 0.041(6) & 1.69
 & 0.692(4) & 0.040(7) & 1.71
 \\
 & & $\Upsilon$(3S) 
 & 0.796(13) & 0.053(16) & 0.59
 & 0.794(13) & 0.054(27) & 0.61
 & 0.795(13) & 0.049(17) & 1.15
 & 0.793(13) & 0.050(27) & 1.26
\\
\cline{3-15} 
 & \multirow{3}{*}{163.4} & $\Upsilon$(1S) 
 & 0.515(1) & 0.000(6) & 0.55
 & 0.515(1) & 0.003(6) & 0.60
 & 0.515(1) & 0.008(8) & 0.83
 & 0.515(1) & 0.004(5) & 1.89
\\
 & & $\Upsilon$(2S) 
 & 0.689(4) & 0.024(10) & 0.92
 & 0.689(4) & 0.027(11) & 0.91
 & 0.690(4) & 0.027(9)  & 1.50
 & 0.688(3) & 0.023(11) & 1.71
 \\
 & & $\Upsilon$(3S) 
 & 0.803(16) & 0.047(17) & 0.63
 & 0.803(15) & 0.052(23) & 0.65
 & 0.792(14) & 0.038(20) & 0.98
 & 0.800(1)  & 0.037(22) & 1.72
\\
\hline\hline
&&&
\multicolumn{6}{c}{Cut Lorentzian uncorrelated fits with cut=$4\Gamma_0$} &
\multicolumn{6}{c}{Cut Lorentzian correlated fits with cut=$4\Gamma_0$} &      
\\  
& & & \multicolumn{3}{c}{Excl. $\delta$ term} &
      \multicolumn{3}{c}{Incl. $\delta$ term} &
      \multicolumn{3}{c}{Excl. $\delta$ term} &
      \multicolumn{3}{c}{Incl. $\delta$ term} 
\\
\cline{4-6} 
\cline{7-9} 
\cline{10-12} 
\cline{13-15} 
&&
&$aM_{\text{med}}$ & $a\sqrt{c_2}$ & \tiny $\frac{\chi^2}{\text{d.o.f.}}$
&$aM_{\text{med}}$ & $a\sqrt{c_2}$ & \tiny $\frac{\chi^2}{\text{d.o.f.}}$
&$aM_{\text{med}}$ & $a\sqrt{c_2}$ & \tiny $\frac{\chi^2}{\text{d.o.f.}}$
&$aM_{\text{med}}$ & $a\sqrt{c_2}$ & \tiny $\frac{\chi^2}{\text{d.o.f.}}$ 
\\
\hline \hline
\vspace{-3ex}
\multirow{9}{*}{0.0602}  & & & & & &
\\
 & \multirow{3}{*}{204.9} & $\Upsilon$(1S) 
 & 0.515(1) & 0.026(4) & 0.56
 & 0.515(1) & 0.015(2) & 0.78
 & 0.515(1) & 0.028(5) & 0.98
 & 0.515(1) & 0.015(4) & 1.72
\\
 & & $\Upsilon$(2S) 
 & 0.693(4) & 0.086(7) & 1.53
 & 0.691(4) & 0.058(5) & 1.56
 & 0.694(4) & 0.090(7) & 2.06
 & 0.692(3) & 0.060(7) & 2.46
\\
 & & $\Upsilon$(3S) 
 & 0.801(13) & 0.128(13) & 0.86
 & 0.797(15) & 0.084(13) & 1.02
 & 0.802(13) & 0.130(13) & 1.28
 & 0.800(14) & 0.089(9) & 1.27
\\
\cline{3-15} 
 & \multirow{3}{*}{182.1} & $\Upsilon$(1S) 
 & 0.516(1) & 0.018(7) & 0.61
 & 0.516(1) & 0.012(4) & 0.55
 & 0.517(1) & 0.025(5) & 0.95
 & 0.516(1) & 0.012(4) & 1.24
 \\
 & & $\Upsilon$(2S) 
 & 0.692(4) & 0.056(11) & 1.16
 & 0.691(3) & 0.037(7) & 1.12
 & 0.694(4) & 0.063(10) & 1.69
 & 0.691(3) & 0.039(6) & 1.64
 \\
 & & $\Upsilon$(3S) 
 & 0.795(12) & 0.081(22) & 0.58
 & 0.793(12) & 0.053(23) & 0.61
 & 0.794(12) & 0.076(25) & 1.17
 & 0.793(12) & 0.050(20) & 1.33
\\
\cline{3-15} 
 & \multirow{3}{*}{163.4} & $\Upsilon$(1S) 
 & 0.515(1) & 0.000(9) & 0.55
 & 0.515(1) & 0.004(6) & 0.60
 & 0.515(1) & 0.013(13) & 0.83
 & 0.515(1) & 0.003(6) & 2.17
\\
 & & $\Upsilon$(2S) 
 & 0.689(4) & 0.037(16) & 0.92
 & 0.689(4) & 0.026(11) & 0.91
 & 0.690(4) & 0.042(14) & 1.51
 & 0.688(3) & 0.024(11) & 1.62
 \\
 & & $\Upsilon$(3S) 
 & 0.803(15) & 0.073(26) & 0.63
 & 0.802(14) & 0.051(20) & 0.66
 & 0.792(13) & 0.044(26) & 0.98
 & 0.792(13) & 0.026(22) & 1.29
\\
\hline\hline
\end{tabular}
}
\caption{Summary of parameters from correlated and uncorrelated fits on continuum-subtracted correlators or effective masses using cut Lorentzian ans\"atze, taking $\Upsilon$ case as an example.}
\label{tab:fit-Lorentzian-spacing0602-parameters-Uspilon}
\end{table}

\begin{table}[htbp]
\centering
\resizebox{\textwidth}{!}{
\begin{tabular}{*{17}{c}}
\hline\hline
\multirow{3}{*}{$a$}                                &
\multirow{3}{*}{$T$}                                &
\multirow{3}{*}{State}                              &
\multicolumn{6}{c}{Gaussian-type uncorrelated fits} &
\multicolumn{6}{c}{Gaussian-type correlated fits}   &          
\\  
& & & \multicolumn{3}{c}{Excl. $\delta$ term} &
      \multicolumn{3}{c}{Incl. $\delta$ term} &
      \multicolumn{3}{c}{Excl. $\delta$ term} &
      \multicolumn{3}{c}{Incl. $\delta$ term} 
\\
\cline{4-6} 
\cline{7-9} 
\cline{10-12} 
\cline{13-15} 
[fm] & [MeV] &
& $aM_{\text{med}}$ & $a\sqrt{c_2}$ & \tiny $\frac{\chi^2}{\text{d.o.f.}}$
& $aM_{\text{med}}$ & $a\sqrt{c_2}$ & \tiny $\frac{\chi^2}{\text{d.o.f.}}$
& $aM_{\text{med}}$ & $a\sqrt{c_2}$ & \tiny $\frac{\chi^2}{\text{d.o.f.}}$
& $aM_{\text{med}}$ & $a\sqrt{c_2}$ & \tiny $\frac{\chi^2}{\text{d.o.f.}}$ 
\\
\hline \hline 
\vspace{-3ex}
\multirow{15}{*}{0.0493} & & & & & & &
\\
 & \multirow{3}{*}{249.8} & $\Upsilon$(1S) 
 & 0.614(1) & 0.024(1) & 0.37
 & 0.614(1) & 0.022(5) & 0.46
 & 0.614(1) & 0.023(2) & 1.72
 & 0.618(4) & 0.032(8) & 1.61
\\
 & & $\Upsilon$(2S) 
 & 0.760(3) & 0.075(3) & 0.87
 & 0.755(4) & 0.064(8) & 0.55
 & 0.763(3) & 0.073(2) & 1.14 
 & 0.758(5) & 0.063(9) & 0.76
\\
 & & $\Upsilon$(3S) 
 & 0.843(9) & 0.093(8) & 0.17 
 & 0.841(9) & 0.089(11) & 0.38 
 & 0.847(9) & 0.092(9) & 1.73 
 & 0.846(11) & 0.089(17) & 1.10
\\
\cline{3-15} 
 & \multirow{3}{*}{222.2} & $\Upsilon$(1S) 
 & 0.613(1) & 0.015(1) & 0.31 
 & 0.613(1) & 0.015(1) & 0.27 
 & 0.613(1) & 0.014(2) & 1.76 
 & 0.613(1) & 0.014(5) & 1.45
\\
 & & $\Upsilon$(2S) 
 & 0.757(3) & 0.059(3) & 0.39
 & 0.756(3) & 0.055(6) & 0.34 
 & 0.759(3) & 0.058(4) & 1.87 
 & 0.759(4) & 0.052(14) & 1.75
\\
 & & $\Upsilon$(3S)
 & 0.845(9) & 0.083(7) & 0.17
 & 0.842(11) & 0.073(12) & 0.57 
 & 0.842(8) & 0.071(9) & 1.44 
 & 0.836(9) & 0.053(17) & 1.09
\\
\cline{3-15} 
 & \multirow{3}{*}{200.0} & $\Upsilon$(1S) 
 & 0.614(1) & 0.011(3) & 0.16
 & 0.614(1) & 0.010(4) & 0.16
 & 0.614(1) & 0.012(3) & 1.28
 & 0.614(1) & 0.012(1) & 1.09
\\
 & & $\Upsilon$(2S)
 & 0.760(5) & 0.048(5) & 0.24
 & 0.761(5) & 0.038(6) & 0.45
 & 0.763(6) & 0.051(6) & 2.09
 & 0.760(3) & 0.038(4) & 1.22
\\
 & & $\Upsilon$(3S) 
 & 0.847(10) & 0.068(9) & 0.75 
 & 0.864(19) & 0.075(19) & 0.43 
 & 0.853(10) & 0.070(9) & 1.37 
 & 0.867(10) & 0.076(7) & 1.41
\\
\cline{3-15} 
 & \multirow{3}{*}{181.8} & $\Upsilon$(1S) 
 & 0.615(1) & 0.010(1) & 0.81
 & 0.615(1) & 0.009(1) & 0.89
 & 0.615(1) & 0.010(1) & 1.23
 & 0.615(1) & 0.010(1) & 0.94
\\
 & & $\Upsilon$(2S) 
 & 0.757(3) & 0.033(5) & 0.97
 & 0.755(3) & 0.027(6) & 0.98
 & 0.758(3) & 0.034(4) & 1.03
 & 0.758(1) & 0.034(2) & 1.73
\\
 & & $\Upsilon$(3S) 
 & 0.844(8) & 0.049(8) & 0.47
 & 0.841(9) & 0.038(18) & 0.41
 & 0.846(8) & 0.050(7) & 1.49
 & 0.842(8) & 0.035(15) & 0.66
\\
\cline{3-15} 
 & \multirow{3}{*}{166.6} & $\Upsilon$(1S)
 & 0.614(1) & 0.005(5) & 1.18
 & 0.614(1) & 0.002(2) & 1.21
 & 0.614(1) & 0.000(1) & 1.22
 & 0.614(1) & 0.000(1) & 1.07
\\
 & & $\Upsilon$(2S) 
 & 0.758(3) & 0.022(7) & 0.78
 & 0.758(3) & 0.024(15) & 0.84
 & 0.757(2) & 0.018(8) & 1.06
 & 0.757(2) & 0.019(6) & 1.09
\\
 & & $\Upsilon$(3S) 
 & 0.845(9) & 0.031(17) & 0.89
 & 0.846(6) & 0.032(1) & 0.78
 & 0.846(8) & 0.031(14) & 1.13
 & 0.849(6) & 0.032(1) & 0.92
\\
\hline \hline
\vspace{-3ex}
\multirow{9}{*}{0.0602}  & & & & & &
\\
 & \multirow{3}{*}{204.9} & $\Upsilon$(1S) 
 & 0.515(1) & 0.019(5) & 0.22
 & 0.514(1) & 0.014(6) & 0.24
 & 0.515(1) & 0.020(6) & 1.11
 & 0.514(1) & 0.016(5) & 1.48
\\
 & & $\Upsilon$(2S) 
 & 0.694(6) & 0.064(6) & 1.13
 & 0.692(5) & 0.061(5) & 1.12
 & 0.696(5) & 0.065(5) & 2.09
 & 0.694(4) & 0.063(3) & 2.24
\\
 & & $\Upsilon$(3S) 
 & 0.803(15) & 0.096(10) & 0.52
 & 0.800(14) & 0.091(12) & 0.69
 & 0.806(15) & 0.096(10) & 1.35
 & 0.804(14) & 0.092(10) & 1.23
\\
\cline{3-15} 
 & \multirow{3}{*}{182.1} & $\Upsilon$(1S) 
 & 0.516(1) & 0.014(5) & 0.23
 & 0.516(1) & 0.013(4) & 0.19
 & 0.517(1) & 0.018(4) & 0.97
 & 0.516(1) & 0.014(3) & 1.08
 \\
 & & $\Upsilon$(2S) 
 & 0.695(7) & 0.045(10) & 0.85
 & 0.693(6) & 0.042(9)  & 0.81
 & 0.696(6) & 0.048(8)  & 1.87
 & 0.695(5) & 0.045(6)  & 1.64
 \\
 & & $\Upsilon$(3S) 
 & 0.802(16) & 0.068(14) & 0.37
 & 0.800(15) & 0.063(16) & 0.45
 & 0.798(15) & 0.059(16) & 1.20
 & 0.799(16) & 0.059(21) & 1.20
\\
\cline{3-15} 
 & \multirow{3}{*}{163.4} & $\Upsilon$(1S)
 & 0.515(1) & 0.000(9) & 0.25
 & 0.515(1) & 0.003(3) & 0.18
 & 0.515(1) & 0.000(6) & 1.29
 & 0.515(1) & 0.003(2) & 1.09
\\
 & & $\Upsilon$(2S) 
 & 0.691(5) & 0.033(10) & 0.69
 & 0.691(5) & 0.031(10) & 0.62
 & 0.691(5) & 0.032(11) & 1.71
 & 0.690(4) & 0.029(6)  & 1.41
 \\
 & & $\Upsilon$(3S)
 & 0.799(13) & 0.045(18) & 0.45
 & 0.794(11) & 0.042(14) & 0.53
 & 0.794(11) & 0.032(17) & 1.56
 & 0.795(11) & 0.032(16) & 1.05
\\
\hline\hline
\end{tabular}
}
\caption{Summary of parameters from correlated and uncorrelated fits on continuum-subtracted correlators or effective masses using Gaussian-type ans\"atze, taking $\Upsilon$ case as an example.}
\label{tab:fit-Gaussian-parameters-Uspilon}
\end{table}

\FloatBarrier

\bibliographystyle{JHEP}
\bibliography{ref.bib}

\end{document}